\documentclass[aps,pre,a4paper,twocolumn,twoside,
unsortedaddress,eqsecnum,showpacs,showkeys]{revtex4}
\usepackage{amsmath,amssymb}
\usepackage[mathscr]{euscript}
\usepackage[bbgreekl]{mathbbol}
\usepackage{pifont}
\usepackage{graphicx}
\allowdisplaybreaks[4]
\newcommand*{\be}{\begin{equation}}
\newcommand*{\ee}{\end{equation}}
\newcommand*{\ba}{\begin{array}}
\newcommand*{\ea}{\end{array}}
\newcommand*{\bea}{\begin{eqnarray}}
\newcommand*{\eea}{\end{eqnarray}}
\newcommand*{\bean}{\begin{eqnarray*}}
\newcommand*{\eean}{\end{eqnarray*}}

\newcommand*{\lp}{\left(}
\newcommand*{\rp}{\right)}
\newcommand*{\ls}{\left[}
\newcommand*{\rs}{\right]}
\newcommand*{\lc}{\left\{}
\newcommand*{\rc}{\right\}}

\newcommand*{\la}{\langle}
\newcommand*{\La}{\left\la}
\newcommand*{\ra}{\rangle}
\newcommand*{\Ra}{\right\ra}

\renewcommand*{\d}{\textrm{d}}
\newcommand*{\e}{\mathrm{e}}
\newcommand*{\g}{\text{\slshape g}}
\newcommand*{\p}{\partial}
\newcommand*{\vphi}{\varphi}

\newcommand*{\kB}{k_{\mathrm{B}}}
\newcommand*{\Tr}{\mathop{{\rm Tr}}\nolimits\,}

\newcommand*{\ds}{\displaystyle}

\newlength{\glength}
\renewcommand*{\g}{\makebox[\glength][c]{\text{\slshape g}}}

\def\tens#1{{\text{\sffamily\bfseries#1}}}

\newcommand*{\CPL}{Chem. Phys. Lett.\ }
\newcommand*{\JACS}{J. Am. Chem. Soc.\ }

\newcommand*{\JCP}{J. Chem. Phys.\ }
\newcommand*{\JML}{J. Mol. Liq.\ }

\newcommand*{\JPC}{J. Phys. Chem.\ }

\newcommand*{\JPSJ}{J. Phys. Soc. Jpn.\ }
\newcommand*{\MP}{Mol. Phys.\ }
\newcommand*{\PRev}{Phys. Rev.\ }
\newcommand*{\PRA}{Phys. Rev. A\ }
\newcommand*{\PRE}{Phys. Rev. E\ }
\newcommand*{\PR}{Phys. Rep.\ }
\newcommand*{\PTP}{Prog. Theor. Phys.\ }
\newcommand*{\RMoP}{Rev. Mod. Phys.\ }

\newcommand*{\scE}{{\text{\textsc e}}}
\newcommand*{\scH}{{\text{\textsc h}}}
\newcommand*{\scL}{{\text{\textsc l}}}

\newcommand*{\shCP}{C_{\scriptstyle P}}
\newcommand*{\CV}{C_{\scriptstyle V}}
\newcommand*{\cP}{c_{\scriptscriptstyle P}}
\newcommand*{\cV}{c_{\scriptscriptstyle V}}
\newcommand*{\vS}{v_{\scriptscriptstyle S}}

\newcommand*{\chiS}{\chi_{\scriptscriptstyle S}}
\newcommand*{\chiT}{\chi_{\scriptscriptstyle T}}

\newcommand*{\rhoC}{\rho_{\text{\textsc c}}}

\newcommand*{\rhoH}{\rho_{\text{\textsc h}}}
\newcommand*{\rhoM}{\rho_{\text{\textsc m}}}

\newcommand*{\drhoH}{\dot{\rho}_{\text{\textsc h}}}
\newcommand*{\drhoM}{\dot{\rho}_{\text{\textsc m}}}

\newcommand*{\hrho}{\hat{\rho}}

\newcommand*{\hrhoE}{\hat{\rho}_{\text{\textsc e}}}
\newcommand*{\hrhoH}{\hat{\rho}_{\text{\textsc h}}}

\newcommand*{\dhrho}{\dot{\hat{\rho}}}

\newcommand*{\dhrhoE}{\dot{\hat{\rho}}_{\text{\textsc e}}}

\newcommand*{\cind}[6]{{\text{\scriptsize%
\setlength{\arraycolsep}{0pt}%
$\ba{cc}#1&#2\\ [-0.3ex](#3)&(#4)\\ [-0.8ex]#5&#6\ea$}}}
\newcommand*{\find}[3]{{\text{\scriptsize%
\setlength{\arraycolsep}{0pt}%
$\ba{c}#1\\ [-0.3ex](#2)\\ [-0.8ex]#3\ea$}}}

\DeclareSymbolFont{rsfs}{U}{rsfs}{m}{n}
\DeclareSymbolFontAlphabet{\mathrsfs}{rsfs}

\begin{document}
\title{Statistical-mechanical theory of
ultrasonic absorption in molecular liquids}

\author{Alexander E. Kobryn}
\affiliation{Department of Theoretical Study,
Institute for Molecular Science, Myodaiji, Okazaki, Aichi 444-8585, Japan}

\author{Fumio Hirata}
\affiliation{Department of Theoretical
Study, Institute for Molecular Science, and Department of Functional
Molecular Science, The Graduate University for Advanced Studies (SOKENDAI),
Myodaiji, Okazaki, Aichi 444-8585, Japan}

\date{October 10, 2006}

\begin{abstract}
We present results of theoretical description of ultrasonic
phenomena in molecular liquids. In particular, we are interested in
the development of microscopical, i.e., statistical-mechanical
framework capable to explain the long living puzzle of the excess
ultrasonic absorption in liquids. Typically, ultrasonic wave in a
liquid can be generated by applying the periodically alternating
external pressure with the angular frequency that corresponds to the
ultrasound. If the perturbation introduced by such process is weak
-- its statistical-mechanical treatment can be done with the use of
the linear response theory. We treat the liquid as a system of
interacting sites, so that all the response/aftereffect functions as
well as the energy dissipation and generalized (wave-vector and
frequency dependent) ultrasonic absorption coefficient are obtained
in terms of familiar site-site static and time correlation functions
such as static structure factors or intermediate scattering
functions. To express the site-site intermediate scattering
functions we refer to the site-site memory equations in the
mode-coupling approximation for the first-order memory kernels,
while equilibrium properties such as site-site static structure
factors, direct and total correlation functions are deduced from the
integral equation theory of molecular liquids known as RISM or one
of its generalizations. All the formalism is phrased in a general
manner, hence the obtained results are expected to work for
arbitrary type of molecular liquid including simple, ionic, polar,
and non-polar liquids.
\end{abstract}

\pacs{%
05.20.Jj -- Statistical Mechanics of Classical Fluids,
61.25.Em -- Molecular Liquids}

\keywords{%
Ultrasonic Absorption Coefficient,
Integral Equation Theory,
Memory Equation,
}

\maketitle

\section{Introduction}
\label{Section01}

Sound velocity and absorption are important probes of physical and
chemical processes taking place in liquid and liquid mixtures
\cite{Herzfeld59,Bhatia67}. Sound velocity reflects density
fluctuations occurring in solutions, thereby, it is closely related
to the compressibility, which is a response function of the density
to the mechanical perturbation, or pressure. The property is
probably the best measure for the mechanical stability of solutions.
On the other hand, sound absorption probes the energy dissipation
caused by a variety of irreversible processes taking place in
solutions: for example, structural relaxation of liquids,
conformational transition of molecules in solution, and so forth. No
wonder why people have employed the method for long time to
investigate the structure and dynamics in solutions. There has been
enormous amount of experimental data accumulated in the filed of
science and technology. More importantly, the method seems to be
finding new horizon in different fields of scientific research
\cite{Shutilov88,Povey97,Cheeke02,Dukhin02,Kundu04}, not mentioning
about the usage in medicine and pharmacy
\cite{Hill04,Rumack05,Attwood81,Tolley83,Tata93,Ishihara93,Jeffers95,Tachibana98,Curra03,Wu05}.
Especially important among the applications is that to the stability
and conformational transition of biomolecules such as protein. The
sound velocity and the adiabatic compressibility have already been
on duty to clarifying the mechanical stability of protein in
solution \cite{Gavish83,Gekko79,Gekko89,Gekko91}. At the frontier in
the science, sound velocity has been employed to identify the
conformational transition from native to denatured states of protein
(for the effect of ultrasound on biological molecules and
macromolecules see, e.g., Refs.
\cite{Choi86,Choi87,Bae96,Sakai00,Pethrick83,Pavlovskaya92,Kharakoz89,%
Kharakoz91,Kharakoz93,Ravichandran91,Shin94,Chalikian95,Kitamura95,Karabutov98}).
It won't be so long before the sound absorption associated with the
conformational relaxation is determined by the acoustic measurement.
The method seems appropriate to investigate the large scale
fluctuation or the collective dynamics characteristic to the
macromolecule, because the frequency range of such motion is covered
by that of sound wave. However, there exists a high barrier for the
method to be overcome in order to be applied to the field. That is
the method or theory to analyze the experimental observations.

Traditional ways of interpreting data from acoustic measurements are
based on the hydrodynamic and/or the phenomenological theories of
relaxation \cite{Landau84,Richards39,Markham51,Bauer49,Gierer50a,%
Teubner79,Bhattacharjee81,Narasimham90,Martynov01,Delgado05}
exemplified by that of Debye-type and of stretched-exponential-type.
Those theories inevitably require phenomenological and/or molecular
models for the relaxation, which sometimes are far from reality. One
of the most successful theories of the sound absorption is that for
water, proposed by Hall more than half a century ago
\cite{Hall47,Hall48}. Hall has assumed that water is an equilibrium
mixture of two states: monomers having densely packed structure and
hydrogen-bonded clusters which have an ice-like open-packed
structure. The equilibrium will shift toward the monomers by
acoustic pressure due to the difference in the molar volume between
the two components of the mixture. The energy dissipation associated
with the relaxation process is probed by the sound absorption. Hall
applied the transition state theory to the two-state model in order
to explain the sound absorption by liquid water. The treatment
should be remarked as the most important contemporary achievement in
theories of sound absorption by liquids, because it is the first
theory to relate the acoustic process with the microscopic model of
the liquid structure.

Unfortunately, the progress seems stopped at the level of Hall as
far as the microscopic treatment of the sound absorption is
concerned with. The reason is simply because Hall's theory employs a
structural model for the liquid. It means that the theory requires a
model for liquid structure for each liquid or solution, and
phenomenological or empirical parameters associated with the model,
such as the molar volume. The difficulty of constructing a model for
liquid structure will be easily understood by taking the
water-alcohol mixture as an example. All those are suggesting that
the theory of sound absorption is necessary to start from a
Hamiltonian level or a molecular interaction, not from structural
models of liquid.

One may think it will then be appropriate to employ the molecular
dynamics simulation, because the method starts definitely from the
Hamiltonian model. However, the problem is formidably difficult for
the method, because it touches both the hydrodynamic and
thermodynamic limits. Just imagine the typical wave length of the
ultrasonic wave, and how many molecules should be involved in a wave
packet. But if the simulation is technically feasible, it does not
help much, because the analysis of trajectory will require the
statistical mechanics of molecular liquids. Without having a method
of analysis, the trajectory is just a garbage.

The present paper aims to construct a molecular theory of sound
absorption based on the latest development in the statistical
mechanics, or the site-site representation of the generalized
Langevin theory. The theory combines two elements in the theoretical
physics: the reference interaction site model (RISM) and the
generalized Langevin equation (GLE). The RISM theory is an integral
equation method to describe equilibrium structure of molecular
liquids in terms of the site-site pair correlation functions, from
which all the thermodynamic quantities can be derived
\cite{Hirata81,Hirata82,Hirata83,Hirata03}. The theory and its three
dimensional generalization have been applied successfully to almost
entire spectrum of chemical and physical processes ta\-king place in
solution from chemical reactions to the molecular recognition by
protein \cite{Hirata03,Imai05,Yoshida06}. Among the applications,
the results for the isothermal compressibility of the water-butanol
mixture \cite{Omelyan03} are of particular interest in conjunction
with the present topics, since it is closely related to the
mechanical stability of solutions. The theory could have
successfully reproduced the concentration and temperature dependence
of the compressibility of the solution observed experimentally, and
it could have been able to describe the behavior in terms of the
liquid structure. In that sense, the theory has already demonstrated
its capability to explain the molecular mechanism of sound velocity.
However, it is not complete for describing the entire physics of
sound propagation, because it is in general accompanied by energy
dissipation, or attenuation of sound wave. This is the reason why we
employ the GLE theory -- a statistical mechanics theory for
irreversible processes. In past two decades, we have developed a
method which combines RISM and GLE \cite{Hirata03}. The method could
have described a variety of irreversible processes with great
success, including the ion transport \cite{Chong98}, dynamics of
water \cite{Chong99a,Chong99b}, viscosity \cite{Yamaguchi01},
dielectric relaxation \cite{Yamaguchi03a}, translational and
rotational dynamics of a molecule in solutions
\cite{Yamaguchi04b,Kobryn05,Kobryn06}. Especially important in
relation with the present topic is the contribution made by
Yamaguchi \textit{et~al.} \cite{Yamaguchi03b} who formulated a
theory for ultrasonic vibration potential, or the coupling of
acoustic and electrostatic perturbations in polar liquids. This work
provides a good guide for the development of a theory of ultrasonic
phenomena in liquids.

In the present paper, we propose a theory for sound absorption in
molecular liquids based on combination of the RISM theory, the GLE
theory, and the linear response theory. Starting from definition of
the perturbation Hamiltonian which describes the coupling of the
sound wave and a liquid system, we derive an expression for the
sound absorption in the most general form. In the choice of
parameters for abbreviated description we recognize that the
traditionally employed set of fluctuations of local site densities
and longitudinal site currents is not enough to serve our purposes
in this case and has to be extended by adding fluctuations of the
total energy density. The latter is really necessary if one wants to
obtain in the expression for ultrasonic absorption coefficient terms
related to the thermal conductivity of the system. We then apply the
theory to the simplest case of a liquid of spherical molecules and
derive expressions in the hydrodynamic limit in order to make
contact with the result from classical hydrodynamics.

The paper is organized as follows. In Section~\ref{Section02} we
spe\-ci\-fy the system, define basic dynamical quantities that we
use for its description, and write their equations of motion. In
Section~\ref{Section03} we provide derivation of the perturbation
Hamiltonian for the case of not too strong ultrasonic vibrations so
that the result of derivation can be used in the theory of linear
response. Application of the theory of linear response to our system
is described in Section~\ref{Section04}, where we calculate linear
responses of local particle number, charge and mass densities for
the perturbation caused by the propagating ultrasonic wave. All
responses are written in terms of familiar static and time
correlation functions. A major step to calculate ultrasonic
absorption coefficient is made in Section~\ref{Section05}, where we
first derive an expression for the energy dissipation, and then use
it to define the \textit{generalized}, i.e., wave-vector and
frequency dependent ultrasonic absorption coefficient in a liquid.
For that purpose one may require the knowledge of site-site
intermediate scattering functions. Therefore another and rather big
portion of the material is dedicated to the problem of their
calculation and is put in Sections~\ref{Section06}
and~\ref{Section07}. The remaining part of the paper is about
detailed analysis of our findings. In particular, we consider the
hydrodynamic limit of our result for the case of one-component
simple liquid in order to compare it with the one from the
continuous media models \cite{Herzfeld59,Bhatia67,Landau84}. The
required steps and their consequences are described in
Section~\ref{Section08}. Finally, advances, deficiencies, remedies
and open questions of our treatment are summarized in
Section~\ref{Section09}. Auxiliary material is put into Appendices.

\section{Definition of the system and variables}
\label{Section02}

\newcommand*{\jone}{\mathbf{j}\hspace{-0.22em}\raisebox{-0.36ex}{1}}
\newcommand*{\jonez}{j\hspace{-0.42em}\raisebox{-0.44ex}{\itshape{1}}}
\newcommand*{\sumone}{\sum_{\text{\ding{172}}}}
\newcommand*{\sumtwo}{\sum_{\text{\ding{173}}}}
\newcommand*{\sumonetwo}{\sum_{\text{\ding{172},\ding{173}}}}

Let us give first generic definitions of major quantities for a
mixture of $\mathcal{M}$-component classical molecular liquid
confined in volume $V$. We suppose that each component consists of
$N_\mu$ particles (and indices $\mu,\nu=1,\ldots,\mathcal{M}$ are
used to label molecular species), hence the total number of
particles is $\sum_{\mu}N_\mu=N$. According to this,
$x_\mu=N_\mu/N$ should be particle relative concentrations (or
molar fractions). Each molecule consists of $\Lambda_\mu$ sites
(could be atoms) having charges $z_\mu^\alpha$ and masses
$m_\mu^\alpha$ (and indices $\alpha,\gamma=1,\ldots,\Lambda_\mu$
are used to label molecular sites). Time-dependent positions and
momenta of individual sites will be denoted then as
$\mathbf{r}_{\mu,i}^{\alpha}(t)$ and
$\mathbf{p}_{\mu,i}^{\alpha}(t)=m_{\mu}^{\alpha}\mathbf{v}_{\mu,i}^{\alpha}(t)$,
respectively, with $\mathbf{v}_{\mu,i}^{\alpha}(t)$ being site
velocities.

Let us suppose that the unperturbed system in concern is described
by the Hamiltonian $H$, and that the perturbation introduced by the
external pressure $P$ is $H'$. Microscopic expression for $H$ can be
written as the sum of the kinetic $H_{\text{kin}}$ and interaction
$H_{\text{int}}$ parts as
\begin{subequations}
\label{eq.2.1}
\bea
H&=&H_{\text{kin}}+H_{\text{int}},\label{eq.2.1a}\\
H_{\text{kin}}&=&\sum_{\mu=1}^{\mathcal{M}}\sum_{i=1}^{N_\mu}\sum_{\alpha=1}^{\Lambda_\mu}
\frac{\ls p_{\mu,i}^\alpha(t)\rs^2}{2m_\mu^\alpha},\label{eq.2.1b}\\
H_{\text{int}}&=&\sum_{\mu=1}^{\mathcal{M}}\sum_{i=1}^{N_\mu}\sum_{\alpha=1}^{\Lambda_\mu}
\sum_{\nu=1}^{\mathcal{M}}\sum_{j=1}^{N_\nu}\sum_{\gamma=1}^{\Lambda_\nu}
\phi_{\mu\nu,ij}^{\alpha\gamma}\lp|\mathbf{r}_{\mu\nu,ij}^{\alpha\gamma}(t)|\rp,
\qquad\label{eq.2.1c}
\eea
\end{subequations}
where $\mathbf{r}_{\mu\nu,ij}^{\alpha\gamma}(t)\equiv
\mathbf{r}_{\mu,i}^\alpha(t)-\mathbf{r}_{\nu,j}^\gamma(t)$, and
$\phi_{\mu\nu,ij}^{\alpha\gamma}(|\mathbf{r}_{\mu\nu,ij}^{\alpha\gamma}(t)|)$
is potential of interaction between sites, also it is assumed
implicitly that if $\nu=\mu$, then $j\not=i$ (in the following,
this condition will be indicated by the prime mark next to the
summation symbol). At present, a detailed knowledge of
structure for $\phi_{\mu\nu,ij}^{\alpha\gamma}
(|\mathbf{r}_{\mu\nu,ij}^{\alpha\gamma}(t)|)$ is not required. For
simplicity of notations we will denote the triplicate sum by a
symbol as follows
\bea
\sum_{\mu=1}^{\mathcal{M}}\sum_{i=1}^{N_\mu}\sum_{\alpha=1}^{\Lambda_\mu}
\longrightarrow\sumone,\quad
\sum_{\nu=1}^{\mathcal{M}}\sum_{j=1}^{N_\nu}\sum_{\gamma=1}^{\Lambda_\nu}
\longrightarrow\sumtwo.
\label{eq.2.2}
\eea
Equations of motion for individual site positions and momenta are the
Hamilton equations
\begin{subequations}
\label{eq.2.3}
\bea
\dot{\mathbf{r}}_{\mu,i}^\alpha(t)&=&\frac{\p H}{\p\mathbf{p}_{\mu,i}^\alpha(t)},\label{eq.2.3a}\\
\dot{\mathbf{p}}_{\mu,i}^\alpha(t)&=&-\frac{\p H}{\p\mathbf{r}_{\mu,i}^\alpha(t)}.\label{eq.2.3b}
\eea
\end{subequations}
They are used to derive microscopic equations of motion for
dynamical variables. Microscopic single-particle density and
momentum density are defined, respectively, as
\begin{subequations}
\label{eq.2.4}
\bea
\rho_{\mu,i}^\alpha(\mathbf{r};t)&=&
\delta\lp\mathbf{r}-\mathbf{r}_{\mu,i}^\alpha(t)\rp,\label{eq.2.4a}\\
\mathbf{p}_{\mu,i}^\alpha(\mathbf{r};t)&=&
\mathbf{p}_{\mu,i}^\alpha(t)\delta\lp\mathbf{r}-\mathbf{r}_{\mu,i}^\alpha(t)\rp,
\label{eq.2.4b}
\eea
\end{subequations}
with $\delta(\mathbf{r})$ being the 3d Dirac $\delta$-function. Next
we introduce collective variables such as particle number-density
$\rho(\mathbf{r};t)$, particle charge-density $\rhoC(\mathbf{r};t)$
and particle mass-density $\rhoM(\mathbf{r};t)$, for which we will
use common designation $\bbrho(\mathbf{r};t)$, so that
\bea
\bbrho(\mathbf{r};t)
&=&\Big\{\rho(\mathbf{r};t),\rhoC(\mathbf{r};t),\rhoM(\mathbf{r};t)\Big\}\nonumber\\
&=&\sumone\rho_{\mu,i}^\alpha(\mathbf{r};t)\lc1,z_\mu^\alpha,m_\mu^\alpha\rc,\nonumber\\
&=&\sumone\rho_{\mu,i}^\alpha(\mathbf{r};t)\Upsilon_\mu^\alpha,\label{eq.2.5}
\eea
where $\Upsilon_\mu^\alpha$ is used to denote the set
$\{1,z_\mu^\alpha,m_\mu^\alpha\}$. For the system in equilibrium
$\la\rho(\mathbf{r};t)\ra=\rho$, $\la\rhoC(\mathbf{r};t)\ra=0$ and
$\la\rhoM(\mathbf{r};t)\ra=\rhoM$, where $\rho$ is the mean value of
the particle number-density and $\rhoM$ is unperturbed mass-density.
Angular brackets $\la\ldots\ra$ denote an appropriate statistical
average: for example, the Gibbs canonical ensemble ave\-ra\-ge.

Similarly, the particle number-density current
$\mathbf{j}(\mathbf{r};t)$, charge-density current
$\mathbf{j}_{\text{\sc c}}(\mathbf{r};t)$, and mass-density current
$\mathbf{j}_{\text{\sc m}}(\mathbf{r};t)$, for which we also will
use one common designation $\mathbb{j}(\mathbf{r};t)$, are defined
as
\bea
\mathbb{j}(\mathbf{r};t)&=&
\Big\{\mathbf{j}(\mathbf{r};t),\mathbf{j}_{\text{\sc c}}(\mathbf{r};t),
\mathbf{j}_{\text{\sc m}}(\mathbf{r};t)\Big\}\nonumber\\
&=&\sumone\mathbf{p}_{\mu,i}^\alpha(\mathbf{r};t)
\frac1{m_\mu^\alpha}\Upsilon_\mu^\alpha.\label{eq.2.6}
\eea
It is worth to note, that $\mathbf{j}_{\text{\sc m}}(\mathbf{r};t)$
is nothing but the momentum-density, i.e.,
$\mathbf{j}_{\text{\sc{m}}}(\mathbf{r};t)
\equiv\mathbf{p}(\mathbf{r};t)$. Its spacial integral is the total
momentum of the system $\mathbf{P}$, which is the conserved
quantity. By a suitable choice of the reference frame it can always
be maid to vanish. In the following, we will assume that for the
case of unperturbed system that condition is satisfied.

Energy density $\rhoH(\mathbf{r};t)$ can be introduced in a
similar way, i.e., that its spatial integral is the total system
energy, or the Hamiltonian $H$, and is therefore the conserved
quantity. It is convenient to distinguish explicitly contributions
from kinetic and interaction parts, so that the microscopic
expression for $\rhoH(\mathbf{r};t)$ reads \cite{Zubarev74}
\begin{subequations}
\label{eq.2.7}
\bea
\rhoH(\mathbf{r};t)&=&\rho_{\text{{\sc h},kin}}(\mathbf{r};t)
+\rho_{\text{{\sc h},int}}(\mathbf{r};t),\label{eq.2.7a}\\
\rho_{\text{{\sc h},kin}}(\mathbf{r};t)&=&\sumone
\frac{\ls p_{\mu,i}^\alpha(t)\rs^2}{2m_\mu^\alpha}
\rho_{\mu,i}^\alpha(\mathbf{r};t),\label{eq.2.7b}\\
\rho_{\text{{\sc h},int}}(\mathbf{r};t)&=&\frac12\int\d\mathbf{r}'\;
\sideset{}{'}{\mathop{\sumonetwo}}
\phi_{\mu\nu,ij}^{\alpha\gamma}(|\mathbf{r}-\mathbf{r}'|)
f_{\mu\nu,ij}^{\alpha\gamma}(\mathbf{r},\mathbf{r}';t),
\nonumber\\\label{eq.2.7c}
\eea
\end{subequations}
where $f_{\mu\nu,ij}^{\alpha\gamma}(\mathbf{r},\mathbf{r}';t)$ is
a two-particle distribution function. Its general expression
depends on many factors of the system and can be obtained, in
principle, from the kinetic theory. In our consideration, however,
we will use the approximation
\begin{eqnarray}
\rho_{\text{{\sc h},int}}(\mathbf{r};t)&\approx&
\frac12\int\d\mathbf{r}'\sideset{}{'}{\mathop{\sumonetwo}}
\phi_{\mu\nu,ij}^{\alpha\gamma}(|\mathbf{r}-\mathbf{r}'|)
\g_{\mu\nu,ij}^{\alpha\gamma}(|\mathbf{r}-\mathbf{r}'|)\nonumber\\
&&{}\times\delta\lp\mathbf{r}-\mathbf{r}_{\mu,i}^{\alpha}(t)\rp
\delta\lp\mathbf{r}'-\mathbf{r}_{\nu,j}^{\gamma}(t)\rp,\label{eq.2.8}
\end{eqnarray}
where $\g_{\mu\nu,ij}^{\alpha\gamma}(|\mathbf{r}-\mathbf{r}'|)$ is
the radial distribution function that depends only on the mutual
distance between sites.

Equations of motion for dynamical variables $\bbrho(\mathbf{r};t)$,
$\mathbb{j}(\mathbf{r};t)$ and $\rho_{\text{\sc h}}(\mathbf{r};t)$
can be obtained with the use of Hamilton equations (\ref{eq.2.3}).
In the reciprocal space they take the form of algebraic equations,
since $\mathbf{k}$-dependent quantities $\bbrho(\mathbf{k};t)$,
$\mathbb{j}(\mathbf{k};t)$ and $\rho_{\text{\sc h}}(\mathbf{k};t)$
are spatial 3d-Fo\-u\-rier trans\-forms of their counterparts in the
direct space (we use the same notations in both direct and
reciprocal spaces and hope there will be no confusion since we
distinguish functions by their arguments). In particular, equations
of motion for quantities in concern read
\begin{subequations}
\label{eq.2.9}
\begin{eqnarray}
\drhoM(\mathbf{k};t)&=&i\mathbf{k}\cdot\mathbf{p}(\mathbf{k};t),\label{eq.2.9a}\\
\dot{\mathbf{p}}(\mathbf{k};t)&=&i\mathbf{k}:\mathbf{\Pi}(\mathbf{k};t),\label{eq.2.9b}\\
\drhoH(\mathbf{k};t)&=&i\mathbf{k}\cdot\mathbf{j}_{\text{\sc h}}(\mathbf{k};t)
+ih(\mathbf{k};t),\label{eq.2.9c}
\end{eqnarray}
\end{subequations}
where $\mathbf{\Pi}(\mathbf{k};t)$ is the Fourier transform of the
stress tensor, and $\mathbf{j}_{\text{\sc h}}(\mathbf{k};t)$ and
$h(\mathbf{k};t)$ are related with the Fourier transform of the
energy current (its kinetic and interaction parts, respectively).
For the moment, a detailed knowledge of their structure is not
needed.

\section{The Perturbation Hamiltonian}
\label{Section03}

\begin{figure}[!htb]
\begin{center}
\fbox{\includegraphics*[bb=107 277 488 564,width=0.8\columnwidth]{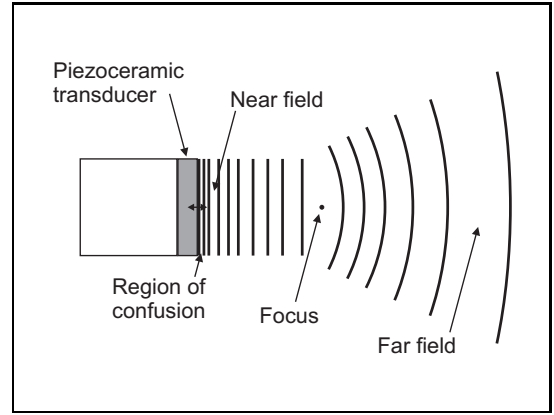}}
\end{center}
\caption{Piston source operating with the diameter (aper\-tu\-re)
very much greater than the acoustic wavelength in the medium in
contact with the front of the transducer \protect\cite{Povey97}. The
sound field in front of a piston source is drawn as lines of a
constant phase, i.e., the positions of maximum or a minimum in a wave.}
\label{Fig.1}
\end{figure}

In this section we are concerned about derivation of the
perturbation Hamiltonian. Hence, it is worth to begin from
considering the generation of ultrasonic waves and their propagation
in a media. A typical type of the ultrasound source in action is
displayed in Fig.~\ref{Fig.1}. The sound field in front of a source
is represented by means of lines of a constant phase, where the
phase of a wave refers to the position of maximum or minimum in a
wave. Experimental observations tell \cite{Herzfeld59,Bhatia67} that
ultrasonic waves lose their intensity with penetration depth
rapidly. It means that sonic amplitude in area called the far field
is negligible. By considering a situation when the piston source
oscillations produce the ultrasonic wave of relatively small
amplitude, one may expect that the theory of linear response
\cite{Kubo57a,Kubo57b} can be applied to describe the introduced
perturbation almost everywhere in front of a piston starting from
the near field and spreading forward to the area beyond the focus.

When the external pressure is applied to the system, the perturbed
part $H'$ of the total Hamiltonian $H_{\text{tot}}$ can be related
to the work on changing of the deformation tensor $u_{\beta\delta}$
(here indices $\beta$, $\delta$ are used to denote Cartesian
coordinates $x$, $y$, $z$). In the linear approximation this tensor
is defined as \cite{Shutilov88}
\bea
u_{\beta\delta}=\frac12\lp\frac{\p l_\beta}{\p r_\delta}
+\frac{\p l_\delta}{\p r_\beta}\rp,\label{eq.3.1}
\eea
where $\mathbf{l}$ is the time-dependent displacement due to the
periodically applied external pressure (which for definiteness we
will think is applied along the $z$-axis of the system), and
therefore should be presented as
\bea
\mathbf{l}\to\mathbf{l}(t)=\mathbf{a}\exp\lc-i\lp\omega t+\vphi\rp\rc\label{eq.3.2}
\eea
with $a=|\mathbf{a}|$, $\omega$ and $\vphi$ being the amplitude,
angular fre\-qu\-en\-cy, and initial phase, respectively. Assuming
that the displacement $\mathbf{l}(t)$ is sufficiently small (in the
macroscopic scale), elementary work on changing the deformation
tensor can be calculated as the internal stress force times the
displacement. Total work is then obtained by integration over the
volume of the system:
\be
\int\d V\;W(\mathbf{r};t)=
\int\d V\sum_{\beta,\delta}
\frac{\p\ls\boldsymbol{\Pi}(\mathbf{r};t)\rs_{\beta\delta}}
{\p\ls\mathbf{r}\rs_\delta}\ls\mathbf{l}(t)\rs_\beta,
\label{eq.3.3}
\ee
where $W(\mathbf{r};t)$ denotes the work of internal stress forces
in unit volume. The density of this work is the density of the
energy of elastic deformation and can be regarded as the density of
the perturbation Hamiltonian $H'(\mathbf{r};t)$. For the following
consideration it is convenient to rewrite it as
\begin{subequations}
\label{eq.3.4}
\bea
H'(\mathbf{r};t)=-M(\mathbf{r};t)E(t),\label{eq.3.4a}
\eea
where
\bea
M(\mathbf{r};t)&=&-\sum_{\beta,\delta}
\frac{\p\ls\boldsymbol{\Pi}(\mathbf{r};t)\rs_{\beta\delta}}
{\p\ls\mathbf{r}\rs_\delta}\ls\mathbf{a}\rs_\beta\label{eq.3.4b}\\
E(t)&=&\exp\lc-i\lp\omega t+\vphi\rp\rc.\label{eq.3.4c}
\eea
\end{subequations}
The reason for that will be seen later.

\section{Linear response\protect\linebreak to the acoustic perturbation}
\label{Section04}

In the theory of linear response \cite{Kubo57a,Kubo57b} the total
system Hamiltonian $H_{\text{tot}}$ is considered to be a sum of its
regular part, say $H$, and perturbation, say $H'$. If the density of
the perturbed part can be written as
$H'(\mathbf{r};t)=-M(\mathbf{r};t)E(t)$, then in terms of the
spatial Fourier components the time-dependent average linear
response of a dynamical quantity $A(t)$ produced by the applied
field $E(t)$ is given by
\be
\la A(\mathbf{k};t)\ra_{\text{\o}}=-\frac1{\kB T}\int_{-\infty}^t\d s
\la \dot{A}(\mathbf{k};t-s)M(-\mathbf{k};0)\ra E(s),\label{eq.4.1}
\ee
where $\kB$ is the Boltzmann constant, $T$ is thermodynamic
temperature, and the subscript \o{} is used to indicate the average
of a \textit{perturbed} quantity. An example: $E$ is an electric
field and $M$ is the electric dipole moment. Both $M$ and $E$ terms
for our case are defined in the previous section.

By imposing fluctuation notations,
$\hat{\bbrho}(\mathbf{k};t)=\bbrho(\mathbf{k};t)-\bbrho$, and taking
$A(\mathbf{k};t)=\hat{\bbrho}(\mathbf{k};t)$, one obtains
\begin{multline}
\la\hat{\bbrho}(\mathbf{k};t)\ra_{\text{\o}}=
\frac{\e^{-i\lp\omega t+\vphi\rp}}{\kB T}\\
{}\times\int_0^\infty\d\tau\;
\e^{i\omega\tau}\sum_{\beta,\delta}ik_\delta a_\beta
\La\dot{\hat{\bbrho}}(\mathbf{k};\tau)
\ls\boldsymbol{\Pi}^*(\mathbf{k};0)\rs_{\beta\delta}\Ra,
\label{eq.4.2}
\end{multline}
where $*$ means complex conjugation. Integrand in the equation above
can be transformed with the use of properties of time-correlation
functions \cite{Harp70,Berne70,Berne76,Steele87} and equations of motion
(\ref{eq.2.9a}) and (\ref{eq.2.9b}). The result can be written
compactly as
\bea
\la\hat{\bbrho}(\mathbf{k};t)\ra_{\text{\o}}=
\frac{ia_k}{\kB T}\;\e^{-i\lp\omega t+\vphi\rp}D_\bbrho(\mathbf{k};\omega)
\label{eq.4.3}
\eea
with $a_k=a/k$ and
\begin{multline}
D_\bbrho(\mathbf{k};\omega)=\sum_{\mu\nu,\alpha\gamma}
m_\mu^\alpha\Upsilon_\nu^\gamma\lp N_\mu N_\nu\rp^{\frac12}\\
\times\lc k^2\ls\tens{J}_{\mu\nu}^{\alpha\gamma}(k)\rs_{zz}
+\omega^2S_{\mu\nu}^{\alpha\gamma}(k)
+i\omega^3\tilde{F}_{\mu\nu}^{\alpha\gamma}(k;\omega)\rc,\label{eq.4.4}
\end{multline}
where $S_{\mu\nu}^{\alpha\gamma}(k)$ is the site-site static
structure factor, $\tilde{F}_{\mu\nu}^{\alpha\gamma}(k;\omega)$ is
the up\-per-half Fourier transform of the site-site intermediate
scattering function $F_{\mu\nu}^{\alpha\gamma}(\mathbf{k};t)$ with
respect to time $t$, and $\tens{J}_{\mu\nu}^{\alpha\gamma}(k)$ is
tensor of site-site static current correlation functions, all of
them introduced in the standard way:
\begin{subequations}
\label{eq.4.5}
\bea
S_{\mu\nu}^{\alpha\gamma}(k)&=&\lp N_\mu N_\nu\rp^{-\frac12}\sum_{i,j}
\la\hrho_{\nu,j}^{\gamma,*}(\mathbf{k};0)\hrho_{\mu,i}^\alpha(\mathbf{k};0)\ra,
\label{eq.4.5a}\\
F_{\mu\nu}^{\alpha\gamma}(k;\tau)&=&\lp N_\mu N_\nu\rp^{-\frac12}\sum_{i,j}
\la\hrho_{\nu,j}^{\gamma,*}(\mathbf{k};0)\hrho_{\mu,i}^\alpha(\mathbf{k};\tau)\ra,
\label{eq.4.5b}\\
\ls\tens{J}_{\mu\nu}^{\alpha\gamma}(k)\rs_{\beta\delta}&=&\lp N_\mu N_\nu\rp^{-\frac12}\sum_{i,j}
\la\ls\mathbf{j}_{\nu,j}^{\gamma,*}(\mathbf{k})\rs_\delta
\ls\mathbf{j}_{\mu,i}^\alpha(\mathbf{k})\rs_\beta\ra,\qquad
\label{eq.4.5c}
\eea
\end{subequations}
and
\bea
\tilde{F}_{\mu\nu}^{\alpha\gamma}(k;\omega)=
\int_0^\infty\d\tau\;\e^{i\omega\tau}F_{\mu\nu}^{\alpha\gamma}(k;\tau).
\label{eq.4.6}
\eea
We also assume for simplicity that the wave-vector $\mathbf{k}$ is
taken to be parallel to the $z$-axis, i.e., $\mathbf{k}=(0,0,k)$.

The applied field to the system (in our case the applied pressure)
must in general be real, so that the full monochromatic ``force''
$E(t)$, Eq. (\ref{eq.3.4c}), should be the superposition (as well as
the displacement (\ref{eq.3.2}))
\bea
\frac12\lp\e^{-i\omega t}+\e^{i\omega t}\rp\equiv\cos\lp\omega t+\vphi\rp,
\label{eq.4.7}
\eea
and the total response function is the superposition of responses
from each component \cite{Kubo57a,Kubo57b,Berne70,Berne76}. It is
quite straightforward then to show that
\begin{multline}
\la\hat{\bbrho}(\mathbf{k};t)\ra_{\text{\o,total}}\\
=\frac{ia_k}{2\kB T}
\lc\e^{-i\lp\omega t+\vphi\rp}D_\bbrho(k;\omega)
+\e^{i\lp\omega t+\vphi\rp}D_\bbrho(k;-\omega)\rc.
\label{eq.4.8}
\end{multline}
In the following, we
will not use the subscript ``total'' next to the symbol \o{} by
assuming that the quantity in concern indicates already the
superposition of responses as has been explained above.

\section{Energy dissipation}
\label{Section05}

The response of the real system to the external
mo\-no\-chro\-ma\-tic perturbation is accompanied by the absorption
and propagation of energy. The reason for that is because under the
influence of the external perturbation the system changes its state
\cite{Berne70}. The difference between the energy absorbed and
emitted is the energy dissipation. The energy $Q(\mathbf{k};\omega)$
dissipated in the unit of time is the averaged over one period of
monochromatic field the time rate of change of the system energy,
and for our case can be written as
\bea
Q(\mathbf{k};\omega)=\frac{\omega}{2\pi}
\int_0^{\frac{2\pi}{\omega}}\d t
\La M(\mathbf{k};t)\Ra_{\text{\o}}\frac{\p E(t)}{\p t}\label{eq.5.1}
\eea
with $M$ and $E$ having their meaning given by de\-fi\-ni\-ti\-ons
(\ref{eq.3.4}). In terms of the linear response it produces
\begin{multline}
Q(\mathbf{k};\omega)=-\frac{\omega^2}{4\kB T}\\
\times\int_0^\infty\d\tau\lc\e^{i\omega\tau}+\e^{-i\omega\tau}\rc
\la M^*(\mathbf{k};0)M(\mathbf{k};\tau)\ra,
\label{eq.5.2}
\end{multline}
where exponential factors are left being not reduced as in
(\ref{eq.4.7}) intensionally in order to show the contribution from
each term. The expression above is calculated as
\bea
Q(\mathbf{k};\omega)&=&
\frac{a_k^2\omega^6}{4\kB T}\sum_{\mu\nu,\alpha\gamma}
m_\mu^\alpha m_\nu^\gamma(N_\mu N_\nu)^{\frac12}\nonumber\\
&&\times
\lc\tilde{F}_{\mu\nu}^{\alpha\gamma}(\mathbf{k},\omega)
+\tilde{F}_{\mu\nu}^{\alpha\gamma}(\mathbf{k},-\omega)\rc,
\label{eq.5.3}
\eea
which means that the energy dissipation in the linear response
theory for a system like our is determined solely   by the site-site
intermediate scattering functions of the system alone.

When the sound wave is propagated through the liquid, its intensity
decreases with the distance. In par\-ti\-cu\-lar, the existence of
viscosity and thermal conductivity results in the dissipation of
energy in sound waves, and the sound is consequently absorbed. The
decrease will occur according to a law $\e^{-2\alpha r}$, and the
amplitude will decrease as $\e^{-\alpha r}$, where the absorption
coefficient $\alpha$ is defined in terms of rate of energy
dissipation $Q$ and the density of the energy of the sound wave
$\bar{Q}$ as \cite{Landau84}
\bea
\alpha=|Q|\Big/2c\bar{Q},\label{eq.5.4}
\eea
where $c$ is the (adiabatic) velocity of sound in the liquid, and
$\bar{Q}$ is given by $\bar{Q}=\frac12\rhoM v_0^2V$ with $v_0$ being
velocity of a fluid in the sound wave. Following the definition, let
us introduce generalized (wave-vector and frequency dependent)
absorption coefficient $\alpha(\mathbf{k};\omega)$ by the relation
\bea
\alpha(\mathbf{k};\omega)=
\frac{|Q(\mathbf{k};\omega)|}{2\vS(\mathbf{k})\bar{Q}(\omega)},
\label{eq.5.5}
\eea
where $\vS(\mathbf{k})$ denotes $\mathbf{k}$-dependent (adiabatic)
sound velocity, and we use $\bar{Q}\to\bar{Q}(\omega)=\frac12\rhoM
v_0^2(\omega)V$, in which
$\mathbf{v}_0\to\mathbf{v}_0(\omega)=\max\ls-\mathbf{a}\,
\omega\sin\lp\omega t+\vphi\rp\rs=\mathbf{a}\,\omega$ is the
velocity of fluid due to the external perturbation and not due to
the thermal motion. When $Q(\mathbf{k};\omega)$ and
$\bar{Q}(\omega)$ are substituted into (\ref{eq.5.5}) the result is
\bea
\alpha(\mathbf{k};\omega)&=&
\frac{\omega^4}{4k^2\kB T\rhoM\vS(\mathbf{k})}\sum_{\mu\nu,\alpha\gamma}
m_\mu^\alpha m_\nu^\gamma(\rho_\mu \rho_\nu)^{\frac12}\nonumber\\
&&\times
\lc\tilde{F}_{\mu\nu}^{\alpha\gamma}(\mathbf{k},\omega)
+\tilde{F}_{\mu\nu}^{\alpha\gamma}(\mathbf{k},-\omega)\rc.
\label{eq.5.6}
\eea
This is the first ever expression for ultrasonic absorption
coefficient in liquid obtained from the microscopic, i.e.,
statistical-mechanical theory. Its frequency dependence is
determined by the product of $\omega^4$ and site-site dynamic
structure factors
$\tilde{F}_{\mu\nu}^{\alpha\gamma}(\mathbf{k},\pm\omega)$. Also the
result itself is independent of oscillation amplitude $a$ as it
should be.

Since the amplitude characteristics are related by linear relations,
the exponential law of attenuation with generalized absorption
coefficient (which also can be called as the generalized amplitude
attenuation or generalized spatial attenuation coefficient)
$\alpha(\mathbf{k};\omega)$ is valid for any related acoustic
parameter. For example, the generalized damping time coefficient
that characterizes the damping of the wave in time is introduced as
$\tau(\mathbf{k};\omega)=\ls\vS(\mathbf{k})\alpha(\mathbf{k};\omega)\rs^{-1}$,
so that the generalized temporal attenuation coefficient becomes
$\theta(\mathbf{k};\omega)=\ls\tau(\mathbf{k};\omega)\rs^{-1}
=\vS(\mathbf{k})\alpha(\mathbf{k};\omega)$, and the generalized
logarithmic damping decrement becomes
$\vartheta=\mathrsfs{T}\theta(\mathbf{k};\omega)$, where
$\mathrsfs{T}$ is the period of ultrasonic wave.

We see, then, that the detailed analysis of expression for either
$\la\hat{\bbrho}(\mathbf{k};t)\ra_\text{\o}$,
$Q(\mathbf{k};\omega)$, or $\alpha(\mathbf{k};\omega)$ is possible
after evaluation of $D_\bbrho(k;\omega)$. This is considered in the
following sections.

\section{TCF\lowercase{s} and the site-site memory equation}
\label{Section06}

Frequency/time dependence of $D_\bbrho(k;\omega)$, Eq.
(\ref{eq.4.4}), is determined by the term
$m_\mu^\alpha\Upsilon_\nu^\gamma\tilde{F}_{\mu\nu}^{\alpha\gamma}(k;\omega)$,
i.e., by fluctuations in number-, charge-, and mass-densities of the
system. A suitable way to calculate such time-correlation functions
is based on the approach by Mori \cite{Mori65a,Mori65b}. Although we
follow it in this section, we shall not re\-peat the very details of
this method: they can be found either in the original source or in
numerous publications after. But definition and handling of a set of
the so-called slow variables of the system are given, which is
required at least to establish notations.

Let us consider a set of dynamical variables
\bea
\mathfrak{A}_{\scL,\scH}=\Big\{\lc\hrho_\mu^\alpha(\mathbf{k};t)\rc,
\lc j_{\mu,\scL}^\alpha(\mathbf{k};t)\rc,\hrhoH(\mathbf{k};t)\Big\},
\label{eq.6.1}
\eea
i.e., the set consisting of three subsets of fluctuations of site
number-densities, longitudinal site current-densities, and
fluctuation of energy density, respectively. In the following, it
will be desirable to distinguish between these three subsets by
assigning them consecutive labels (1), (2), and (3). From the
time-/space-inversion symmetry properties of elements of the set
$\mathfrak{A}_{\scL,\scH}$ one can see that
$j_{\mu,\scL}^\alpha(\mathbf{k})$ is orthogonal to both
$\hrho_\mu^\alpha(\mathbf{k})$ and $\hrhoH(\mathbf{k})$, while
$\hrho_\mu^\alpha(\mathbf{k})$ and $\hrhoH(\mathbf{k})$ are not
mutually orthogonal. For the sake of simplicity of future
considerations it is advantageous to replace $\hrhoH(\mathbf{k})$
with its renormalized variable $\hrhoE(\mathbf{k})$ defined as
\begin{multline}
\hrhoE(\mathbf{k})=\hrhoH(\mathbf{k})
-\sum_{\mu,\nu;\alpha,\gamma}(N_\mu N_\nu)^{-\frac12}\\
\times\hrho_\mu^\alpha(\mathbf{k})\ls\tens{S}^{-1}(k)\rs_{\mu\nu}^{\alpha\gamma}
\la\hrho_\nu^{\gamma,*}(\mathbf{k})\hrhoH(\mathbf{k})\ra,\label{eq.6.2}
\end{multline}
so that
\bea
\la\hrho_\nu^{\gamma,*}(\mathbf{k})\hrhoE(\mathbf{k})\ra=0.\label{eq.6.3}
\eea
We will denote the new set of dynamical variables as
\bea
\mathfrak{A}_{\scL,\scE}=\Big\{\lc\hrho_\mu^\alpha(\mathbf{k};t)\rc,
\lc j_{\mu,\scL}^\alpha(\mathbf{k};t)\rc,\hrhoE(\mathbf{k};t)\Big\}.\label{eq.6.4}
\eea
The microscopic equations of motion for the elements
of the set $\mathfrak{A}_{\scL,\scE}$ read
\begin{subequations}
\label{eq.6.5}
\bea
\dhrho_\mu^\alpha(\mathbf{k};t)&=&ik\;j_{\mu,\scL}^\alpha(\mathbf{k};t),\label{eq.6.5a}\\
\dot{j}_{\mu,\scL}^\alpha(\mathbf{k};t)&=&ik\;\Pi_{\mu,zz}^\alpha(\mathbf{k};t),\label{eq.6.5b}\\
\dhrhoE(\mathbf{k};t)&=&ik\;j_{\scE,\scL}(\mathbf{k};t)+h(\mathbf{k};t).\label{eq.6.5c}
\eea
\end{subequations}
where $\Pi_{\mu,zz}^\alpha(\mathbf{k};t)$  and
$j_{\scE,\scL}(\mathbf{k};t)$ are defined by
\begin{subequations}
\label{eq.6.6}
\bea
\Pi_{zz}(\mathbf{k};t)&=&
\sum_{\mu,\alpha}m_\mu^\alpha\;\Pi_{\mu,zz}^\alpha(\mathbf{k};t),\label{eq.6.6a}\\
j_{\scE,\scL}(\mathbf{k})&=&j_{\scH,\scL}(\mathbf{k})
-\sum_{\mu,\nu;\alpha,\gamma}\lp N_\mu N_\nu\rp^{-\frac12}
j_{\mu,\scL}^\alpha(\mathbf{k})\nonumber\\
&&\times\ls\tens{S}^{-1}(k)\rs_{\mu\nu}^{\alpha\gamma}
\la\hrho_\nu^{\gamma,*}(\mathbf{k})\hrhoH(\mathbf{k})\ra.\label{eq.6.6b}
\eea
\end{subequations}
One can see now that arbitrary elements from different subsets of
$\mathfrak{A}_{\scL,\scE}$ are initially orthogonal to each other,
while elements within a same subset are not. It implies that the
matrix of initial TCFs has block-diagonal structure
\bea
\tens{C}(k)=
\lp\ba{lll}
\tens{C}_{(1)(1)}(k)&\tens{0}&\tens{0}\\
\tens{0}&\tens{C}_{(2)(2)}(k)&\tens{0}\\
\tens{0}&\tens{0}&\tens{C}_{(3)(3)}(k)\\
\ea\rp,\label{eq.6.7}
\eea
where $\tens{0}$ are zero-matrices, and $\tens{C}_{(i)(i)}(k)$ are
square submatrices constructed exclusively with the use of elements
from $(i)$th subset:
\begin{subequations}
\label{eq.6.8}
\bea
\ls\tens{C}(k)\rs\cind{\alpha}{\gamma}{1}{1}{\mu}{\nu}&=&
\la\hrho_\nu^{\gamma,*}(\mathbf{k})\hrho_\mu^\alpha(\mathbf{k})\ra,\label{eq.6.8a}\\
%=\lp N_\mu N_\nu\rp^{-\frac12}\ls\tens{S}(k)\rs_{\mu\nu}^{\alpha\gamma},\\
\ls\tens{C}(k)\rs\cind{\alpha}{\gamma}{2}{2}{\mu}{\nu}&=&
\la j_{\nu,\scL}^{\gamma,*}(\mathbf{k})j_{\mu,\scL}^\alpha(\mathbf{k})\ra,\label{eq.6.8b}\\
%=\lp N_\mu N_\nu\rp^{-\frac12}\ls\tens{L}_\scL(k)\rs_{\mu\nu}^{\alpha\gamma},\\
\ls\tens{C}(k)\rs\cind{}{}{3}{3}{}{}&=&
\la\hrhoE^*(\mathbf{k})\hrhoE(\mathbf{k})\ra.\label{eq.6.8c}
\eea
\end{subequations}
Important property of the block-diagonal matrix is that its
inverse has similar block-diagonal structure:
\bea
\tens{C}^{-1}(k)=
\lp\ba{lll}
\tens{C}_{(1)(1)}^{-1}(k)&\tens{0}&\tens{0}\\
\tens{0}&\tens{C}_{(2)(2)}^{-1}(k)&\tens{0}\\
\tens{0}&\tens{0}&\tens{C}_{(3)(3)}^{-1}(k)\\
\ea\rp.\label{eq.6.9}
\eea
That feature will be used later. Meanwhile, one has to note that in
the case $t\ne0$ matrix of TCFs is not necessary block-diagonal. For
simplicity of notations, we will denote elements of the entire set
$\mathfrak{A}_{\scL,\scE}$ at time $t$ by $A_\lambda(\mathbf{k};t)$,
where $\lambda$ is a composite label of the element of the set.
Vector-row formed by these elements and vector-column formed by
their complex conjugates will be denoted, respectively, by
$\mathbf{A}(\mathbf{k};t)$ and $\mathbf{A}^*(\mathbf{k};t)$.
Therefore the second-rank tensor of TCFs $\tens{C}(k;t)$ constructed
from elements of the set $\mathfrak{A}_{\scL,\scE}$, and its
components read, respectively,
\begin{subequations}
\label{eq.6.10}
\bea
\tens{C}(k;t)&=&
\la\mathbf{A}^*(\mathbf{k};0)\mathbf{A}(\mathbf{k};t)\ra,\label{eq.6.10a}\\\relax
\ls\tens{C}(k;t)\rs_{\lambda_1\lambda_2}&=&
\la A_{\lambda_2}^*(\mathbf{k};0)A_{\lambda_1}(\mathbf{k};t)\ra.\label{eq.6.10b}
\eea
\end{subequations}
If one has to be more specific, it may be necessary to assign
indices $\lambda$ their values which can be $\mu$, $\alpha$, and
the subset number $(i)$. For example:
\begin{subequations}
\label{eq.6.11}
\bea
\ls\tens{C}(k;t)\rs\cind{\alpha}{\gamma}{2}{1}{\mu}{\nu}&=&
\la\hrho_\nu^{\gamma,*}(\mathbf{k})j_\mu^\alpha(\mathbf{k};t)\ra,\label{eq.6.11a}\\
\ls\tens{C}(k;t)\rs\cind{\alpha}{}{1}{3}{\mu}{}&=&
\la\hrhoE^*(\mathbf{k})\hrho_\mu^\alpha(\mathbf{k};t)\ra,
\quad\text{etc.}\label{eq.6.11b}
\eea
\end{subequations}
Thus the problem of evaluation of $\alpha(\mathbf{k};\omega)$
amounts to finding out the solution for time development of TCFs
$\tens{C}_{(1)(1)}(k;t)$.

By introducing the projection operator $\mathrsfs{P}$ whose action
on arbitrary variable is described as
\bea
\mathrsfs{P}\ldots=\sum_{\lambda_1,\lambda_2}\la A_{\lambda_1}^*(\mathbf{k})\ldots\ra
\ls\tens{C}^{-1}(k)\rs_{\lambda_1\lambda_2}A_{\lambda_2}(\mathbf{k}),\quad\label{eq.6.12}
\eea
and its complementary projection operator
$\mathrsfs{Q}=1-\mathrsfs{P}$, one can obtain the so-called memory
equation for TCFs
$C_{\lambda_1\lambda_2}(k;t)\equiv\ls\tens{C}(k;t)\rs_{\lambda_1\lambda_2}$
in the form \cite{Mori65a,Mori65b,Balucani94,Balucani03}
\begin{multline}
\dot{C}_{\lambda_1\lambda_2}(k;t)=
\sum_{\lambda'}i\Omega_{\lambda_1\lambda'}(k)C_{\lambda'\lambda_2}(k;t)\\
-\sum_{\lambda'}\int_0^t\d\tau\;K_{\lambda_1\lambda'}(k;\tau)C_{\lambda'\lambda_2}(k;t-\tau),
\label{eq.6.13}
\end{multline}
where $i\Omega_{\lambda_1\lambda_2}(k)\equiv\ls
i\boldsymbol{\Omega}(k)\rs_{\lambda_1\lambda_2}$ are elements of the
so-cal\-led frequency matrix given as
\begin{eqnarray}
i\Omega_{\lambda_1\lambda_2}(k)=\sum_{\lambda'}
\la A_{\lambda'}^*(\mathbf{k})i\mathrsfs{L}A_{\lambda_1}(\mathbf{k})\ra
\ls\tens{C}^{-1}(k)\rs_{\lambda'\lambda_2},\qquad\label{eq.6.14}
\end{eqnarray}
$K_{\lambda_1\lambda_2}(k;t)\equiv\ls\tens{K}(k;t)\rs_{\lambda_1\lambda_2}$
are elements of the matrix of the first order memory kernels
introduced as
\begin{eqnarray}
K_{\lambda_1\lambda_2}(k;t)=\sum_{\lambda'}
\la\mathfrak{f}_{\lambda'}^*(\mathbf{k})\mathfrak{f}_{\lambda_1}(\mathbf{k};t)\ra
\ls\tens{C}^{-1}(k)\rs_{\lambda'\lambda_2},\quad\label{eq.6.15}
\end{eqnarray}
and $\mathfrak{f}_\lambda(\mathbf{k};t)$ are the so-called
fluctuating forces defined as
\bea
\mathfrak{f}_\lambda(\mathbf{k};t)&=&
\exp\lc i\mathrsfs{QL}t\rc\mathfrak{f}_\lambda(\mathbf{k};0)\nonumber\\
&=&\e^{i\mathrsfs{QL}t}\ls\dot{A}_\lambda(\mathbf{k})
-\sum_{\lambda'}i\Omega_{\lambda\lambda'}(k)A_{\lambda'}(\mathbf{k})\rs.\qquad
\label{eq.6.16}
\eea
In equations above, $\mathrsfs{L}$ is the Liouville operator of the
system. Explicit structure of all nonzero elements of matrices
(\ref{eq.6.14})--(\ref{eq.6.16}) is listed in
Appendix~\ref{AppendixA}.

Since in our case all dynamical variables are related to specific
interacting sites, we shall call the corresponding memory-equations
(and subsequently all quantities involved) the \textit{site-site}
memory-equations. Study of dynamic processes in molecular liquids
with the use of the site-site memory equations has been under
development in last decade in works by Hirata and coworkers (see,
e.g., Refs.
\cite{Chong98,Chong99a,Chong99b,Yamaguchi01,Yamaguchi03a,Yamaguchi04b,Kobryn05,Kobryn06,Yamaguchi03b}
and references therein). The present treatment is, however,
substantially different, and the difference appears in the choice of
the set of slow variables: fluctuations of the total energy density
are included into the set for the first time. As we shall se later,
it is essential in order to reproduce the expression for the
so-called classical coefficient of ultrasonic absorption in liquids.
Sol\-ving equation (\ref{eq.6.13}) may give (in principle) the
answer about the structure of intermediate scattering functions
$F_{\mu\nu}^{\alpha\gamma}(k;t)$ through the finding
$\tens{C}_{(1)(1)}(k;t)$. Its solution in the frequency domain is
written briefly as
\bea
\tilde{\tens{C}}_{(1)(1)}(k;\omega)=
\tilde{\tens{Y}}{}^{-1}(k;\omega)\tilde{\tens{C}}_{(1)(1)}(k;0),
\label{eq.6.17}
\eea
where $\tilde{\tens{C}}_{(1)(1)}(k;0)\equiv\tens{C}_{(1)(1)}(k)$ is
matrix of initial va\-lu\-es, and
$\tilde{\tens{Y}}{}^{-1}(k;\omega)$ can be regarded as a sort of
propagator since all frequency/time dependence of the solution is
determined by this quantity. Required steps to find the formal
solution to the site-site memory equation as well as the solution
itself as a function of frequency terms and memory kernels is
displayed in Appendix~\ref{AppendixB}. General expression for
$\tilde{\tens{Y}}(k;\omega)$ is rather lenghty, but if time is long
and wave-vector is sufficiently small (in the case of propagation of
ultrasonic wave both these conditions are satisfied), cross memory
terms can be neglected leading~to
\begin{widetext}
\vspace*{-2ex}
\bea
\tilde{\tens{Y}}(k;\omega)&=&\tilde{\tens{K}}_{(1)(1)}(k;\omega)-\tens{I}i\omega
\nonumber\\
&&-i\boldsymbol{\Omega}_{(1)(2)}(k)
\lc\tilde{\tens{K}}_{(2)(2)}(k;\omega)-\tens{I}i\omega
-i\boldsymbol{\Omega}_{(2)(3)}(k)
\ls\tilde{\tens{K}}_{(3)(3)}(k;\omega)-\tens{I}i\omega\rs^{-1}
i\boldsymbol{\Omega}_{(3)(2)}(k)\rc^{-1}
i\boldsymbol{\Omega}_{(2)(1)}(k).
\qquad\label{eq.6.18}
\eea
\end{widetext}
One can also replace in remaining expressions for memory kernels the
evolution operator with complementary projection by an ordinary one,
and at the same time evaluate the results at the leading order
$k^2$. Such operation, however, cannot be justified for the general
case, and it makes sense to talk about the final solution to the
me\-mo\-ry equation if one specifies the model for the me\-mo\-ry
kernels. Since ultrasonic frequency is sufficiently low, the
dynamics of the system should be considered in the long-time limit,
i.e., in which the time scale is large enough for fast relaxation
processes to be completed. Memory kernel of the ordinary me\-mo\-ry
equation in that case is usually constructed by the so-called
mode-coupling approximation \cite{Balucani94,Balucani03}. In works
by Chong \textit{et al.} \cite{Chong98b,Chong02} the conventional
mode-coupling theory has been extended to the case of molecular
li\-qu\-ids based on the interaction-site model. One shall follow
this procedure for the numerical evaluation. Albeit, some analytical
analysis is possible for the case of the one-component simple liquid
in the hydrodynamic limit.

\section{Initial values and frequency terms}
\label{Section07}

In order to be solved, either analytically or numerically, site-site
memory equation (\ref{eq.6.13}) requires information about initial
value. Since we are interested in site-site intermediate scattering
functions, the initial values of our concern are site-site static
structure factors. They can be determined within the frame of the
integral equation theory of liquids called RISM, or one of its
generalizations (e.g., extended RISM
\cite{Hirata03,Hirata81,Hirata82,Hirata83}, \textit{etc.}). It
predicts static structure of molecular fluids via the calculation of
site-site pair correlation functions. This method has been
extensively used and proved to be the powerful tool in the
microscopic description of equilibrium qualities of the system. The
main equation for mixture can be written in the reciprocal space in
matrix notations as
\bea
\boldsymbol{\rho}\tens{h}\boldsymbol{\rho}
=\boldsymbol{\omega}\tens{c}\boldsymbol{\omega}
+\boldsymbol{\omega}\tens{c}\boldsymbol{\rho}\tens{h}\boldsymbol{\rho},
\label{eq.7.1}
\eea
where $\boldsymbol{\rho}$ is diagonal matrix consisting of number
densities of each molecular species, while
$\tens{h}\equiv\tens{h}(k)$, $\tens{c}\equiv\tens{c}(k)$ and
$\boldsymbol{\omega}\equiv\boldsymbol{\omega}(k)$ are matrices of
the Fourier transform of site-site total, direct and intra-molecular
correlation functions, respectively. Equations (\ref{eq.7.1}) are
solved with the closure specified. Typical closures are hyper-netted
chain (HNC), mean-spherical approximation (MSA), Percus-Yevick (PY),
\textit{etc.}~\cite{Hirata03}. Then the matrix of site-site
structure factors is given by
\bea
\tens{S}(k)=\boldsymbol{\rho}^{-1}\boldsymbol{\omega}(k)+\boldsymbol{\rho}\tens{h}(k).
\label{eq.7.2}
\eea

The static site-site current correlation function can be treated
analytically. The expression for arbitrary shape of the molecule has
been given recently \cite{Yamaguchi04b,Kobryn05} and reads
\bea
\lefteqn{\ds\ls\tens{J}_{\text{\sc l}}(k)\rs^{\alpha\gamma}=
\frac{\kB T}{M}j_0(k\ell^{\alpha\gamma})}\nonumber\\
&&{}+\frac{\kB T}3\ls j_0(k\ell^{\alpha\gamma})+j_2(k\ell^{\alpha\gamma})
\rs\nonumber\\
&&\qquad\qquad\times[\delta\mathbf{r}^\alpha]^{\text{\sc t}}
\cdot\ls\lp\Tr\pmb{\mathtt{I}}^{-1}\rp\tens{I}-\pmb{\mathtt{I}}^{-1}\rs
\cdot\delta\mathbf{r}^\gamma\nonumber\\
&&{}-\frac{\kB T}{(\ell^{\alpha\gamma})^2}j_2(k\ell^{\alpha\gamma})
[\delta\mathbf{r}^\alpha\mathop{\text{\footnotesize\ding{54}}}
\delta\mathbf{r}^\gamma]^{\text{\sc t}}\cdot\pmb{\mathtt{I}}^{-1}
\cdot[\delta\mathbf{r}^\alpha\mathop{\text{\footnotesize\ding{54}}}
\delta\mathbf{r}^\gamma],\qquad%\nonumber\\
\label{eq.7.3}
\eea
where $M$ is the total mass of the molecule, $\ell^{\alpha\gamma}$
is the distance between sites, $\delta\mathbf{r}^\alpha$ is the
vector pointed from the center of mass to the site $\alpha$,
$\pmb{\mathtt{I}}$ is matrix of inertia moments of the molecule,
$\tens{I}$ has the same meaning of the diagonal unit matrix;
finally, $j_0$ and $j_2$ are spherical Bessel functions of the first
kind of zero and second order, respectively, and symbol
$\mathop{\text{\footnotesize\ding{54}}}$ is used to denote the outer
vector product.

Expressions (\ref{eq.7.2}) and (\ref{eq.7.3}) are crucial in order
to calculate the solution (\ref{eq.6.17}), linear response terms
(\ref{eq.4.4}), and frequency terms (\ref{eq.6.14}). Actually,
calculation of frequency terms $i\boldsymbol{\Omega}_{(1)(2)}(k)$
and $i\boldsymbol{\Omega}_{(2)(1)}(k)$ is rather trivial, while
calculation of $i\boldsymbol{\Omega}_{(2)(3)}(k)$ and
$i\boldsymbol{\Omega}_{(3)(2)}(k)$ may be difficult. The difficulty
is twofold: one is caused by $b_\mu^\alpha(k)$, and the other by
$\la\hrhoE^*(\mathbf{k})\hrhoE(\mathbf{k})\ra$. In the former case
one has to deal with three-body correlation functions, i.e.,
$\la\hrhoH^*(\mathbf{k})\Pi_{\mu,zz}^\alpha(\mathbf{k})\ra$ and
$\la\hrhoH^*(\mathbf{k})\hrho_\nu^\gamma(\mathbf{k})\ra$, while in
the latter case it is essentially the problem of calculation of
four-body correlation function. It follows from the definition of
the microscopic energy density of the system, eqs. (\ref{eq.2.7})
and (\ref{eq.2.8}), where interaction part involves the two-body
distribution function. There is no reliable mechanism in statistical
mechanics which is capable to handle this accurately for the general
case. Experiments also do not reveal many-body correlations directly
(unlike pair correlations) \cite{Egelstaff92}. If one does not wish
to involve computer simulations, all that leaves quite limited space
to maneuver and narrows numbers of choices to just one:
approximations. Particular approximations may be case dependent, but
fortunately there is one that may work for arbitrary system. It is
given by hydrodynamic limit and for most cases can be calculated
nearly exactly. Therefore approximation of $b_\mu^\alpha(k)$ and
$\la\hrhoE^*(\mathbf{k})\hrhoE(\mathbf{k})\ra$ by their hydrodynamic
values sounds naturally in view of the slowness of ultrasonic
processes and is considered as an effective choice for numerical
evaluations.

\section{Relation with hydrodynamics}
\label{Section08}

In order to demonstrate the robustness of our theory, here we
examine its hydrodynamic limit. For that purpose we confine
ourselves to the case of one-component simple li\-qu\-id. The
expression for the so-called \textit{classical ultrasonic absorption
coefficient in liquids} $\alpha_{\text{\sc cl}}$ derived within the
framework of a continuous media theory is well known
\cite{Herzfeld59,Bhatia67,Landau84}:
\bea
\alpha_{\text{\sc cl}}=\frac{\omega^2}{2\rhoM\vS^3}
\lc\lp\frac43\eta_{\text{\sc s}}+\eta_{\text{\sc v}}\rp
+\lambda_{\text{\sc t}}\lp\frac1{\CV}-\frac1{\shCP}\rp\rc.\quad
\label{eq.8.1}
\eea
Here $\vS$ is (zero frequency) adiabatic sound velocity in the
liquid, $\eta_{\text{\sc s}}$ and $\eta_{\text{\sc v}}$ are
phenomenological coefficients of shear and bulk viscosities,
respectively, $\lambda_{\text{\sc t}}$ is phenomenological
coefficient of its thermal conductivity, and $\CV$ and $\shCP$ are
specific heat capacities per unit mass of the liquid at constant
volume and pressure, respectively.

The starting point to relate our theory and hydrodynamics would be
the formal solution to the equation for intermediate scattering
function specified by eqs. (\ref{eq.6.17}) and (\ref{eq.6.18}) in
which $\tilde{Y}(k;\omega)$ for one-component simple liquid reduces
to
\bea
\tilde{Y}(k;\omega)&=&-i\omega\nonumber\\
&&-\frac{i\Omega_{12}(k)\,i\Omega_{21}(k)}
{\ds-i\omega+\tilde{K}_{22}(k;\omega)-\frac{i\Omega_{23}(k)\,i\Omega_{32}(k)}
{\ds-i\omega+\tilde{K}_{33}(k;\omega)}}.\qquad
\label{eq.8.2}
\eea
Expressions for the frequency terms in that case read
\begin{subequations}
\label{eq.8.3}
\bea
i\Omega_{12}(k)&=&ik,\label{eq.8.3a}\\
i\Omega_{21}(k)&=&ik\frac{\kB T}{mS(k)},\label{eq.8.3b}\\
i\Omega_{23}(k)&=&\frac{ik}{m}\frac{b(k)}
{\la\hrhoE^*(\mathbf{k})\hrhoE(\mathbf{k})\ra},\label{eq.8.3c}\\
i\Omega_{32}(k)&=&\frac{ikb^*(k)}{N\kB T},\label{eq.8.3d}
\eea
\end{subequations}
where $S(k)$ is the static structure factor of the liquid, and
\bea
b(k)=\la\hrhoH^*(\mathbf{k})\Pi_{zz}(\mathbf{k})\ra
-\frac{\kB T}{S(k)}\la\hrhoH^*(\mathbf{k})\hrho(\mathbf{k})\ra.
\label{eq.8.4}
\eea
In the hydrodynamic, i.e., $k\to0$ limit
\bea
S(k)|_{k\to0}&=&\kB T\rho\chiT,\label{eq.8.5}\\
b(k)|_{k\to0}
%&=&\kB T\lp\frac{\p\la\Pi_{zz}(k\to0)\ra}{\p T}\rp_V\nonumber\\
%&=&\kB T^2V\lp\frac{p P}{\p T}\rp_V\nonumber\\
%&=&-\kB T^2V\lp\frac{\p P}{\p V}\rp_T\lp\frac{\p V}{\p T}\rp_P\nonumber\\
&=&\kB T^2V\frac{\beta_{\scriptscriptstyle P}}
{\chi_{\scriptscriptstyle T}},\label{eq.8.6}\\
\la\hrhoE^*(\mathbf{k})\hrhoE(\mathbf{k})\ra|_{k\to0}&=&\kB T^2N\cV,\label{eq.8.7}
\eea
where $\chi_{\scriptscriptstyle S,T}$ and $\beta_{\scriptscriptstyle
P}$ are adiabatic/isothermal compressibility and isobaric thermal
expansion coefficients, respectively, defined in thermodynamics as
\bea
\chi_{\scriptscriptstyle S,T}&=&-\frac1V\lp\frac{\p V}{\p P}\rp_{S,T},\label{eq.8.8}\\
\beta_{\scriptscriptstyle P}&=&\frac1V\lp\frac{\p V}{\p T}\rp_P,\label{eq.8.9}
\eea
and $c_{\scriptscriptstyle P,V}$ is specific heat capacity per
particle at constant pressure/volume. To complete the treatment of
frequency terms one may also need the thermodynamic
de\-fi\-ni\-ti\-on of adiabatic/isothermal sound velocity
$v_{\scriptscriptstyle S,T}$, i.e.,
\bea
v^2_{\scriptscriptstyle S,T}&=&
\lp\frac{\p P}{\p\rhoM}\rp_{S,T}=
\frac1{\rhoM\chi_{\scriptscriptstyle S,T}},
\label{eq.8.10}
\eea
and thermodynamic expression that relates heat capacities at
constant pressure and volume, i.e.,
\bea
\cP-\cV=\frac{T\beta^2_{\scriptscriptstyle P}}{\rho\chi_{\scriptscriptstyle T}}.
\label{eq.8.11}
\eea
With all that in mind, the corresponding products of frequency terms
result in
\begin{subequations}
\label{eq.8.12}
\bea
i\Omega_{12}(k)i\Omega_{21}(k)|_{k\to0}&=&-k^2\frac{\vS^2}{\gamma},\label{eq.8.12a}\\
i\Omega_{23}(k)i\Omega_{32}(k)|_{k\to0}&=&-k^2\vS^2\lp1-\frac1\gamma\rp,\label{eq.8.12b}
\eea
\end{subequations}
where $\gamma=\cP/\cV=\chiT/\chiS$.

Remaining memory kernels $\tilde{K}_{22}(k;\omega)$ and
$\tilde{K}_{22}(k;\omega)$ should be treated as follows. It is known
\cite{Balucani94,Balucani03} that ge\-ne\-ra\-li\-zed longitudinal
viscosity, shear viscosity, and thermal conductivity coefficients,
respectively, can be written in terms of familiar TCFs as
\begin{subequations}
\label{eq.8.13}
\bea
\eta_{\text{\sc l}}(\mathbf{k};t)&\propto&\frac1{\kB TV}
\la\Pi_{zz}^*(\mathbf{k};0)\Pi_{zz}(\mathbf{k};t)\ra,\label{eq.8.13a}\\
\eta_{\text{\sc s}}(\mathbf{k};t)&\propto&\frac1{\kB TV}
\la\Pi_{zx}^*(\mathbf{k};0)\Pi_{zx}(\mathbf{k};t)\ra,\label{eq.8.13b}\\
\lambda_{\text{\sc t}}(\mathbf{k};t)&\propto&\frac1{\kB T^2V}
\la j_{\scE,\scL}^*(\mathbf{k};0)j_{\scE,\scL}(\mathbf{k};t)\ra.\label{eq.8.13c}
\eea
These expressions usually are complemented by the definition of
the generalized bulk viscosity coefficient
$\eta_{\text{\sc v}}(\mathbf{k};t)$ as
\bea
\eta_{\text{\sc l}}(\mathbf{k};t)=
\frac43\eta_{\text{\sc s}}(\mathbf{k};t)
+\eta_{\text{\sc v}}(\mathbf{k};t).\label{eq.8.13d}
\eea
\end{subequations}
The proportionality sign in (\ref{eq.8.13a})--(\ref{eq.8.13c}) can be
replaced by the equality sign after specifying the convention used
to identify the scalar product of two dynamical quantities, say $A$
and $B$, with their time-correlation function. The standard
correspondence $(A,B)=\la A^*B\ra-\la A^*\ra\la B\ra$ is due to Mori
\cite{Mori65a} and has to be used if both $A$ and $B$ have nonzero
averages. This is precisely happening in the case for
$\eta_{\text{\sc l}}(\mathbf{k};t)$, since \cite{Balucani94}
$\la\Pi_{zz}(\mathbf{k})\ra|_{k\to0}=(2\pi)^3\delta(\mathbf{k})\rho\kB
T$. Changing from the continuous to the discrete reciprocal space,
the correct definition for $\eta_{\text{\sc l}}(\mathbf{k};t)$ is
obtained as
\bea
\eta_{\text{\sc l}}(\mathbf{k};t)|_{k\to0}&=&\frac1{\kB TV}
\la[\Pi_{zz}^*(\mathbf{k}\to0)-PV]\nonumber\\
&&\times\e^{i\mathrsfs{L}t}[\Pi_{zz}(\mathbf{k}\to0)-PV]\ra,\label{eq.8.14}
\eea
while both $\eta_{\text{\sc s}}(\mathbf{k};t)$ and
$\lambda_{\text{\sc t}}(\mathbf{k};t)$ do not require correction
since $\la\Pi_{zx}(\mathbf{k})\ra|_{k\to0}=0$ by definition and $\la
j_{\scE,\scL}(\mathbf{k})\ra|_{k\to0}=0$ by the convention used in
this paper (that was discussed in Section~\ref{Section02}).
Re\-la\-ti\-ons of generalized transport co\-ef\-fi\-ci\-ents
$\mathscr{T}(\mathbf{k};t)\equiv\{\eta_{\text{\sc l}}(\mathbf{k};t),
\eta_{\text{\sc s}}(\mathbf{k};t), \eta_{\text{\sc
v}}(\mathbf{k};t), \lambda_{\text{\sc t}}(\mathbf{k};t)\}$ with
ordinary hydrodynamic longitudinal, shear, and bulk viscosities, and
thermal conductivity $\mathscr{T}\equiv\{\eta_{\text{\sc
l}},\eta_{\text{\sc s}}, \eta_{\text{\sc v}},\lambda_{\text{\sc
t}}\}$, respectively, read
\begin{subequations}
\label{eq.8.15}
\bea
\mathscr{T}&=&\int_0^\infty\d t\;\lim_{k\to0}
\mathscr{T}(\mathbf{k};t),\label{eq.8.15a}\\
\eta_{\text{\sc l}}&=&\frac43\eta_{\text{\sc s}}+\eta_{\text{\sc v}}.
\label{eq.8.15b}
\eea
\end{subequations}
Hence, in the hydrodynamic limit memory kernels are identified as
\begin{subequations}
\label{eq.8.16}
\bea
\tilde{K}_{22}(k;\omega)|_{k\to0,\omega\to0}&=&
k^2\nu_{\text{\sc l}},\label{eq.8.16a}\\
\tilde{K}_{33}(k;\omega)|_{k\to0,\omega\to0}&=&
k^2D_{\text{\sc t}}\gamma,\label{eq.8.16b}
\eea
\end{subequations}
where $\nu_{\text{\sc l}}=\eta_{\text{\sc l}}/\rhoM$ is the
longitudinal kinematic viscosity, and $D_{\text{\sc
t}}=\lambda_{\text{\sc t}}/\rho\cP$ is the thermal diffusivity.

When expressions (\ref{eq.8.12}) and (\ref{eq.8.16}) are substituted
into (\ref{eq.8.2}), the solution for $\tilde{F}(\mathbf{k};\omega)$
in the hydrodynamic limit is obtained as
\begin{multline}
\tilde{F}(\mathbf{k};\omega)|_{\text{hyd}}=F(\mathbf{k};0)|_{\text{hyd}}\\
\times\ls-i\omega+\frac{k^2\vS^2/\gamma}
{-i\omega+k^2\nu_{\text{\sc l}}+
\frac{\ds k^2\vS^2\lp1-1/\gamma\rp}
{\ds-i\omega+k^2D_{\text{\sc t}}\gamma}}\relax\rs^{-1},
\label{eq.8.17}
\end{multline}
where $F(\mathbf{k};0)|_{\text{hyd}}\equiv S(k)|_{k\to0}$ and is
given by Eq.~(\ref{eq.8.5}). By using that to calculate
$Q(\mathbf{k};\omega)|_{\text{hyd}}$ and saving the only leading
order contributions, the generalized ultrasonic absorption
coefficient in the hydrodynamic limit is obtained as
\be
\alpha(k;\omega)|_{\text{hyd}}=\frac{\omega^2}{2\rhoM\vS^3}
\lc\eta_\scL+m\lambda_{\text{\sc t}}\lp\frac1{\cV}-\frac1{\cP}\rp\!\rc.
\label{eq.8.18}
\ee
Taking into account the difference in definitions of $\shCP$, $\CV$
and $\cP$, $\cV$, one can see that it completely coincides with
$\alpha_{\text{\sc cl}}$ given by Eq. (\ref{eq.8.1}), which means
that our theory is hydrodynamically consistent. And the importance of
fluctuations of the total energy density $\hrhoH(\mathbf{k};t)$ in
the set of slow variables $\mathfrak{A}_{\scL,\scH}$ now is
revealed: it would be impossible to obtain the classical ultrasonic
absorption coefficient with contribution from thermal processes if
$\hrhoH(\mathbf{k};t)$ is not considered explicitly.

\section{Summary}
\label{Section09}

In this work we presented a statistical-mechanical theory for
treatment of such ultrasonic processes as pro\-pa\-ga\-ti\-on and
absorption of ultrasound in molecular liquids. In particular, we
demonstrated the way to calculate generalized, i.e., wave-vector and
frequency dependent ultrasonic absorption coefficient
$\alpha(\mathbf{k};\omega)$. The suggested description is a
combination of the linear response theory by Kubo
\cite{Kubo57a,Kubo57b} and the memory equation formalism by Mori
\cite{Mori65a,Mori65b} extended to the case of molecular liquids
based on the interaction site model \cite{Hirata03}. Our result is
the first one obtained from the microscopic theory, and reflects all
the general peculiarities of ultrasonic pro\-pa\-ga\-ti\-on and
absorption in liquids known from the phenomenological description.
According to classification by Dukhin and Goetz \cite{Dukhin02},
these peculiarities can be sorted in several different mechanisms.

The \textit{viscous} mechanism is hydrodynamic in nature. It is
related to the shear waves generated by particles or groups of
particles oscillating in the acoustic pressure field. The shear
waves appear due to the difference in densities in the vicinity of
particles and next nearest medium. For example: in terms of
Fig.~\ref{Fig.1}, lines of constant phase may be thought as
positions of maximum compression, while places in between them -- as
positions of dilations. As a result, the liquid layers in the
particle vicinity slide relative to each other, and the
non-stationary sliding motion of the liquid near the particle is
referred to as the shear wave. The viscous mechanism is considered
to be the most important for acoustics since it causes losses of
acoustic energy due to the shear friction. In our work, viscosity
related contribution into absorption coefficients is controlled
mostly by the memory kernel (\ref{eq.A4e}), whose hydrodynamic limit
expression is proportional to the standard time-autocorrelation
function of the stress tensor (\ref{eq.8.14}) and is therefore
associated with generalized longitudinal viscosity.

The \textit{thermal} mechanism is thermodynamic in nature and is
related to the temperature gradients in the vi\-ci\-ni\-ty of
particles and next nearest medium. The temperature gradients are due
to the thermodynamic coupling between pressure and temperature.
There is variety of experiments telling that dissipation of acoustic
energy caused by thermal losses is much smaller than one caused by
the viscous mechanism if it is the case of liquids and solutions,
but not liquid metals \cite{Herzfeld59,Bhatia67}. However it may be
dominant for colloidal systems with soft particles including
emulsion and/or latex droplets \cite{Dukhin02}. Although the latter
systems are not a subject of our consideration, the thermal
mechanism of ultrasonic losses in our description is included
explicitly. One of consequences of this inclusion is the presence of
thermal conductivity coefficient of a liquid in the hydrodynamic
limiting expression of its ultrasonic absorption coefficient. The
required con\-di\-ti\-on for this is the consideration of
fluctuations of the total ener\-gy density (\ref{eq.2.7}) as one of
parameters of abbreviated description forming the set
(\ref{eq.6.1}). It is worth to note that in the limit of small
wave-numbers two major components of ultrasonic absorption --
viscous and thermal -- are additive, eqs. (\ref{eq.8.1}) or
(\ref{eq.8.18}), while in general they cannot be separated from each
other. The concept of a combined mechanistic in nature viscous and
thermodynamic in nature thermal treatment of ultrasonic absorption
process in a liquid can be realized only in microscopic description
of the problem as it was demonstrated in our paper.

The \textit{electrokinetic} mechanism describes interaction of
ultrasound with the double layer of par\-ti\-cles/\-mo\-le\-cu\-les.
Oscillation of the charged particle in the acoustic field leads to
generation of an alternating electric field, and consequently to an
alternating electric current. This mechanism is used in
ele\-ctro-aco\-u\-stic measurements. In the interaction site model
of molecular liquids individual sites are usually atoms or atomic
groups of the molecule and therefore often have an electric charge.
Hence, in our theory most of the types of electrostatic interaction
are included through the system Hamiltonian, and most of the types
of interaction of ultrasound with double layer are taken into
account through the formalism of the linear response. In particular,
the response of charge-density fluctuations is described by
eq.~(\ref{eq.4.8}) and provides the basis for calculation of
ultrasonic vibration potential generated in the system.

The short wave-number limit considered at the end of our paper
demonstrated hydrodynamic consistency of our calculations in a sense
that we were able to accurately reproduce the result for ultrasonic
absorption coefficient known from continuous media theories and
called therefore the classical absorption coefficient. On the other
hand, use of hydrodynamic limit for some frequency terms and the
memory kernels, as described in Sections~\ref{Section07} and
\ref{Section08}, is rather a necessity steamed from the fact that
the requirement of analytical handling of many-body correlation
functions at finite wave-numbers represents a considerable
challenge. For all other quantities one shall be able to use their
microscopic expressions.

We expect our theory of ultrasonic absorption to work for most of
the types of molecular liquids, including simple, non-polar, polar
and ionic. The validity of this expectation has to be tested by
applying the presented formalism to some real systems and evaluating
their ultrasonic absorption coefficients numerically. It could very
well be a subject for another paper.

\acknowledgments
\vspace*{-2ex}

This work is supported in part by the Grant-in-Aid for Scientific Research
on Priority Area of ``Water and Biomolecules'' of the Japanese Ministry of
Education, Culture, Sports, Science and Technology (MONBUKAGAKUSHO).
Authors also thank S.-H. Chong for his fruitful discussions.

\begin{appendix}
\begin{widetext}
\section{Matrix elements of the site-site memory equation}
\label{AppendixA}

Nonzero elements of the site-site frequency matrix
$i\boldsymbol{\Omega}(k)$ read
\begin{subequations}
\label{eq.A1}
\begin{align}
\ls i\boldsymbol{\Omega}(k)\rs\cind{\alpha}{\gamma}{1}{2}{\mu}{\nu}&=\sum_{\nu',\gamma'}
\la{j}_{\nu',\scL}^{\gamma',*}(\mathbf{k})\dhrho_\mu^\alpha(\mathbf{k})\ra
\ls\tens{C}^{-1}(k)\rs\cind{\gamma'}{\gamma}{2}{2}{\nu'}{\nu}&=&
ik\sum_{\nu',\gamma'}(N_{\nu'}N_\nu)^{-\frac12}
\la{j}_{\nu',\scL}^{\gamma',*}(\mathbf{k})j_{\mu,\scL}^\alpha(\mathbf{k})\ra
\ls\tens{J}^{-1}_\scL(k)\rs_{\nu'\nu}^{\gamma'\gamma},\label{eq.A1a}\\
\ls i\boldsymbol{\Omega}(k)\rs\cind{\alpha}{\gamma}{2}{1}{\mu}{\nu}&=\sum_{\nu',\gamma'}
\la\hrho_{\nu'}^{\gamma',*}(\mathbf{k})\dot{j}_{\mu,\scL}^\alpha(\mathbf{k})\ra
\ls\tens{C}^{-1}(k)\rs\cind{\gamma'}{\gamma}{1}{1}{\nu'}{\nu}&=&
ik\sum_{\nu',\gamma'}(N_{\nu'}N_\nu)^{-\frac12}
\la{j}_{\nu',\scL}^{\gamma',*}(\mathbf{k})j_{\mu,\scL}^\alpha(\mathbf{k})\ra
\ls\tens{S}^{-1}(k)\rs_{\nu'\nu}^{\gamma'\gamma},\label{eq.A1b}\\
\ls i\boldsymbol{\Omega}(k)\rs\cind{\alpha}{}{2}{3}{\mu}{}&=
\la\hrhoE^*(\mathbf{k})\dot{j}_{\mu,\scL}^\alpha(\mathbf{k})\ra
\ls\tens{C}^{-1}(k)\rs\cind{}{}{3}{3}{}{}&=&
ik\;b_\mu^\alpha(k)/\la\hrhoE^*(\mathbf{k})\hrhoE(\mathbf{k})\ra,\label{eq.A1c}\\
\ls i\boldsymbol{\Omega}(k)\rs\cind{}{\gamma}{3}{2}{}{\nu}&=\sum_{\nu',\gamma'}
\la{j}_{\nu',\scL}^{\gamma',*}(\mathbf{k})\dhrhoE(\mathbf{k})\ra
\ls\tens{C}^{-1}(k)\rs\cind{\gamma'}{\gamma}{2}{2}{\nu'}{\nu}
&=&ik\sum_{\nu',\gamma'}(N_{\nu'}N_\nu)^{-\frac12}{b_{\nu'}^{\gamma',*}(k)}
\ls\tens{J}^{-1}_\scL(k)\rs_{\nu'\nu}^{\gamma'\gamma},\label{eq.A1d}
\end{align}
\end{subequations}
where
\bea
b_\mu^\alpha(k)=\la\hrhoH^*(\mathbf{k})\Pi_{\mu,zz}^\alpha(\mathbf{k})\ra
-\sum_{\nu_1,\nu_2;\gamma_1,\gamma_2}\lp\frac{N_\mu}{N_{\nu_1}}\rp^{\frac12}
\la\hrhoH^*(\mathbf{k})\hrho_{\nu_1}^{\gamma_1}(\mathbf{k})\ra
\ls\tens{S}^{-1}(k)\rs_{\nu_1\nu_2}^{\gamma_1\gamma_2}
\ls\tens{J}_\scL(\mathbf{k})\rs_{\mu\nu_2}^{\alpha\gamma_2}.\label{eq.A2}
\eea
Nonzero elements of the site-site vector of fluctuating forces
$\tens{f}(\mathbf{k})$ at initial time $t=0$ are then obtained in
the form
\begin{subequations}
\label{eq.A3}
\begin{align}
\ls\tens{f}(\mathbf{k})\rs\find{\alpha}{1}{\mu}&=ik\ls j_{\mu,\scL}^\alpha(\mathbf{k})
-\sum_{\nu,\nu';\gamma,\gamma'}\lp\frac{N_\mu}{N_\nu}\rp^{\frac12}
\ls\tens{J}_\scL(k)\rs_{\mu\nu'}^{\alpha\gamma'}
\ls\tens{J}^{-1}_\scL(k)\rs_{\nu'\nu}^{\gamma'\gamma}
j_{\nu,\scL}^\gamma(\mathbf{k})\rs
&\equiv&ik\;\bar{j}_{\mu,\scL}^\alpha(\mathbf{k}),\label{eq.A3a}\\
\ls\tens{f}(\mathbf{k})\rs\find{\alpha}{2}{\mu}&=ik\ls\Pi_{\mu,zz}^\alpha(\mathbf{k})
-\sum_{\nu,\nu';\gamma,\gamma'}\lp\frac{N_\mu}{N_\nu}\rp^{\frac12}
\ls\tens{J}_\scL(k)\rs_{\mu\nu'}^{\alpha\gamma'}
\ls\tens{S}^{-1}(k)\rs_{\nu'\nu}^{\gamma'\gamma}\hrho_\nu^\gamma(\mathbf{k})
-\frac{b_\mu^\alpha(k)\hrhoE(\mathbf{k})}
{\la\hrhoE^*(\mathbf{k})\hrhoE(\mathbf{k})\ra}\rs
&\equiv&ik\;\bar{\Pi}_{\mu,zz}^\alpha(\mathbf{k}),\label{eq.A3b}\\
\ls\tens{f}(\mathbf{k})\rs\find{}{3}{}&=ik\ls j_{\scE,\scL}(\mathbf{k})
-\sum_{\nu,\nu';\gamma,\gamma'}(N_{\nu'}N_\nu)^{-\frac12}{b_{\nu'}^{\gamma',*}(k)}
\ls\tens{J}^{-1}_\scL(k)\rs_{\nu'\nu}^{\gamma'\gamma}
j_{\nu,\scL}^\gamma(\mathbf{k})\rs+h(\mathbf{k})
&\equiv&ik\;\bar{j}_{\scE,\scL}(\mathbf{k}).\label{eq.A3c}
\end{align}
\end{subequations}
In terms of elements of a set of dynamical variables
$\mathfrak{A}_{\scL,\scE}$, Eq. (\ref{eq.6.4}), site-site frequency
matrix $i\boldsymbol{\Omega}(k)$, eqs. (\ref{eq.A1}), and site-site
vector of fluctuating forces $\tens{f}(\mathbf{k})$, eqs.
(\ref{eq.A3}), elements of the matrix of the first order site-site
memory kernels $\tens{K}(k;t)$ are expresses as
\begin{subequations}
\label{eq.A4}
%\bea
\begin{align}
\ls\tens{K}(k;t)\rs\cind{\alpha}{\gamma}{1}{1}{\mu}{\nu}
&=k^2\sum_{\nu',\gamma'}\lp N_{\nu}N_{\nu'}\rp^{-\frac12}
\la\bar{j}_{\nu',\scL}^{\gamma',*}(\mathbf{k})\e^{i\mathrsfs{QL}t}
\bar{j}_{\mu,\scL}^\alpha(\mathbf{k})\ra
\ls\tens{S}^{-1}(k)\rs_{\nu'\nu}^{\gamma'\gamma},\label{eq.A4a}\\
\ls\tens{K}(k;t)\rs\cind{\alpha}{\gamma}{1}{2}{\mu}{\nu}
&=k^2\sum_{\nu',\gamma'}\lp N_{\nu}N_{\nu'}\rp^{-\frac12}
\la\bar{\Pi}_{\nu',zz}^{\gamma',*}(\mathbf{k})\e^{i\mathrsfs{QL}t}
\bar{j}_{\mu,\scL}^\alpha(\mathbf{k})\ra
\ls\tens{J}^{-1}_\scL(k)\rs_{\nu'\nu}^{\gamma'\gamma},\label{eq.A4b}\\
\ls\tens{K}(k;t)\rs\cind{\alpha}{}{1}{3}{\mu}{}
&=k^2\la\bar{j}_{\scE,\scL}^*(\mathbf{k})\e^{i\mathrsfs{QL}t}
\bar{j}_{\mu,\scL}^\alpha(\mathbf{k})\ra
\Big/\la\hrhoE^*(\mathbf{k})\hrhoE(\mathbf{k})\ra,\label{eq.A4c}\\
\ls\tens{K}(k;t)\rs\cind{\alpha}{\gamma}{2}{1}{\mu}{\nu}
&=k^2\sum_{\nu',\gamma'}\lp N_{\nu}N_{\nu'}\rp^{-\frac12}
\la\bar{j}_{\nu',\scL}^{\gamma',*}(\mathbf{k})\e^{i\mathrsfs{QL}t}
\bar{\Pi}_{\mu,zz}^\alpha(\mathbf{k})\ra
\ls\tens{S}^{-1}(k)\rs_{\nu'\nu}^{\gamma'\gamma},\label{eq.A4d}\\
\ls\tens{K}(k;t)\rs\cind{\alpha}{\gamma}{2}{2}{\mu}{\nu}
&=k^2\sum_{\nu',\gamma'}\lp N_{\nu}N_{\nu'}\rp^{-\frac12}
\la\bar{\Pi}_{\nu',zz}^{\gamma',*}(\mathbf{k})\e^{i\mathrsfs{QL}t}
\bar{\Pi}_{\mu,zz}^\alpha(\mathbf{k})\ra
\ls\tens{J}^{-1}_\scL(k)\rs_{\nu'\nu}^{\gamma'\gamma},\label{eq.A4e}\\
\ls\tens{K}(k;t)\rs\cind{\alpha}{}{2}{3}{\mu}{}
&=k^2\la\bar{j}_{\scE,\scL}^*(\mathbf{k})\e^{i\mathrsfs{QL}t}
\bar{\Pi}_{\mu,zz}^\alpha(\mathbf{k})\ra
\Big/\la\hrhoE^*(\mathbf{k})\hrhoE(\mathbf{k})\ra,\label{eq.A4f}\\
\ls\tens{K}(k;t)\rs\cind{}{\gamma}{3}{1}{}{\nu}
&=k^2\sum_{\nu',\gamma'}\lp N_{\nu}N_{\nu'}\rp^{-\frac12}
\la\bar{j}_{\nu',\scL}^{\gamma',*}(\mathbf{k})\e^{i\mathrsfs{QL}t}
\bar{j}_{\scE,\scL}(\mathbf{k})\ra
\ls\tens{S}^{-1}(k)\rs_{\nu'\nu}^{\gamma'\gamma},\label{eq.A4g}\\
\ls\tens{K}(k;t)\rs\cind{}{\gamma}{3}{2}{}{\nu}
&=k^2\sum_{\nu',\gamma'}\lp N_{\nu}N_{\nu'}\rp^{-\frac12}
\la\bar{\Pi}_{\nu',zz}^{\gamma',*}(\mathbf{k})\e^{i\mathrsfs{QL}t}
\bar{j}_{\scE,\scL}(\mathbf{k})\ra
\ls\tens{J}^{-1}_\scL(k)\rs_{\nu'\nu}^{\gamma'\gamma},\label{eq.A4h}\\
\ls\tens{K}(k;t)\rs\cind{}{}{3}{3}{}{}
&=k^2\la\bar{j}_{\scE,\scL}^*(\mathbf{k})\e^{i\mathrsfs{QL}t}
\bar{j}_{\scE,\scL}(\mathbf{k})\ra
\Big/\la\hrhoE^*(\mathbf{k})\hrhoE(\mathbf{k})\ra.\label{eq.A4i}
\end{align}
\end{subequations}
\end{widetext}

\section{Formal solution\protect\\ to the memory equation}
\label{AppendixB}

\noindent We are interested in finding an expression for
in\-ter\-me\-di\-a\-te scattering functions
$F_{\mu\nu}^{\alpha\gamma}(k;t)$, i.e.,
$\tens{C}_{(1)(1)}(k;t)$ if it is in matrix notations. It is
convenient to do calculations in the reciprocal space, after
applying to the memory equation the upper-half Fourier transform.
There will we three type of equations for
$\tilde{\tens{C}}_{(1)(1)}(k;\omega)$:
\begin{subequations}
\label{eq.B1}
\bea
\lefteqn{-i\omega\tilde{\tens{C}}_{(1)(1)}(k;\omega)=
\tens{C}_{(1)(1)}(k;0)}\nonumber\\
&&+i\boldsymbol{\Omega}_{(1)(2)}(k)\tilde{\tens{C}}_{(2)(1)}(k;\omega)\nonumber\\
&&-\tilde{\tens{K}}_{(1)(1)}(k;\omega)\tilde{\tens{C}}_{(1)(1)}(k;\omega)\nonumber\\
&&-\tilde{\tens{K}}_{(1)(2)}(k;\omega)\tilde{\tens{C}}_{(2)(1)}(k;\omega)\nonumber\\
&&-\tilde{\tens{K}}_{(1)(3)}(k;\omega)\tilde{\tens{C}}_{(3)(1)}(k;\omega),\label{eq.B1a}\\[2ex]
\lefteqn{-i\omega\tilde{\tens{C}}_{(2)(1)}(k;\omega)=
i\boldsymbol{\Omega}_{(2)(1)}(k)\tilde{\tens{C}}_{(1)(1)}(k;\omega)}\nonumber\\
&&+i\boldsymbol{\Omega}_{(2)(3)}(k)\tilde{\tens{C}}_{(3)(1)}(k;\omega)\nonumber\\
&&-\tilde{\tens{K}}_{(2)(1)}(k;\omega)\tilde{\tens{C}}_{(1)(1)}(k;\omega)\nonumber\\
&&-\tilde{\tens{K}}_{(2)(2)}(k;\omega)\tilde{\tens{C}}_{(2)(1)}(k;\omega)\nonumber\\
&&-\tilde{\tens{K}}_{(2)(3)}(k;\omega)\tilde{\tens{C}}_{(3)(1)}(k;\omega),\label{eq.B1b}\\[2ex]
\lefteqn{-i\omega\tilde{\tens{C}}_{(3)(1)}(k;\omega)=
i\boldsymbol{\Omega}_{(3)(2)}(k)\tilde{\tens{C}}_{(2)(1)}(k;\omega)}\nonumber\\
&&-\tilde{\tens{K}}_{(3)(1)}(k;\omega)\tilde{\tens{C}}_{(1)(1)}(k;\omega)\nonumber\\
&&-\tilde{\tens{K}}_{(3)(2)}(k;\omega)\tilde{\tens{C}}_{(2)(1)}(k;\omega)\nonumber\\
&&-\tilde{\tens{K}}_{(3)(3)}(k;\omega)\tilde{\tens{C}}_{(3)(1)}(k;\omega).\label{eq.B1c}
\eea
\end{subequations}
By substituning into (\ref{eq.B1a}) expression for
$\tilde{\tens{C}}_{(2)(1)}(k;\omega)$ from (\ref{eq.B1b}) and
expression for $\tilde{\tens{C}}_{(3)(1)}(k;\omega)$ from
(\ref{eq.B1c}), the formal solution for
$\tilde{\tens{C}}_{(1)(1)}(k;\omega)$ can be written as
\bea
\tilde{\tens{C}}_{(1)(1)}(k;\omega)=
\ls\tilde{\tens{Y}}(k;\omega)\rs^{-1}\tilde{\tens{C}}_{(1)(1)}(k;0),\label{eq.B2}
\eea
where matrix $\tilde{\tens{Y}}(k;\omega)$ has the form
\begin{widetext}
\bea
\lefteqn{\ds\tilde{\tens{Y}}(k;\omega)=\tilde{\tens{K}}_{(1)(1)}(k;\omega)-\tens{I}i\omega
-\tilde{\tens{K}}_{(1)(3)}(k;\omega)
\ls\tilde{\tens{K}}_{(3)(3)}(k;\omega)-\tens{I}i\omega\rs^{-1}
\tilde{\tens{K}}_{(3)(1)}(k;\omega)}\nonumber\\
&&-\lc i\boldsymbol{\Omega}_{(1)(2)}(k)-\tilde{\tens{K}}_{(1)(2)}(k;\omega)
-\tilde{\tens{K}}_{(1)(3)}(k;\omega)
\ls\tilde{\tens{K}}_{(3)(3)}(k;\omega)-\tens{I}i\omega\rs^{-1}
\ls i\boldsymbol{\Omega}_{(3)(2)}(k)-\tilde{\tens{K}}_{(3)(2)}(k;\omega)\rs\rc
\nonumber\\
&&\times\lc\tilde{\tens{K}}_{(2)(2)}(k;\omega)-\tens{I}i\omega
-\ls i\boldsymbol{\Omega}_{(2)(3)}(k)-\tilde{\tens{K}}_{(2)(3)}(k;\omega)\rs
\ls\tilde{\tens{K}}_{(3)(3)}(k;\omega)-\tens{I}i\omega\rs^{-1}
\ls i\boldsymbol{\Omega}_{(3)(2)}(k)-\tilde{\tens{K}}_{(3)(2)}(k;\omega)\rs\rc^{-1}
\nonumber\\
&&\times\lc i\boldsymbol{\Omega}_{(2)(1)}(k)-\tilde{\tens{K}}_{(2)(1)}(k;\omega)
-\ls i\boldsymbol{\Omega}_{(2)(3)}(k)-\tilde{\tens{K}}_{(2)(3)}(k;\omega)\rs
\ls\tilde{\tens{K}}_{(3)(3)}(k;\omega)-\tens{I}i\omega\rs^{-1}
\tilde{\tens{K}}_{(3)(1)}(k;\omega)\rc.\label{eq.B3}
\eea
%
%\clearpage
\end{widetext}

\end{appendix}


\begin{thebibliography}{00}

\bibitem{Herzfeld59}
K. F. Herzfeld and T. A. Litovitz,
\textit{Absorption and Dispersion of Ultrasonic Waves}.
In Pure and Applied Physics: a Series of Monographs and Textbooks, Vol. 7
(Academic Press, New York, 1959).

\bibitem{Bhatia67}
A. B. Bhatia,
\textit{Ultrasonic Absorption: an Introduction to the Theory of Sound Absorption
and Dispersion in Gases, Liquids and Solids}.
In Monographs on the Physics and Chemistry of Materials
(Clarendon Press, Oxford, 1967)
[Reprinted by Dover Publications, New York, 1985].

\bibitem{Shutilov88}
V. A. Shutilov,
\textit{Fundamental Physics of Ultrasound}
(Gordon and Breach, London, 1988).

\bibitem{Povey97}
M. J. W. Povey,
\textit{Ultrasonic Techniques for Fluids Characterization}
(Academic Press, New York, 1997).

\bibitem{Cheeke02}
J. D. N. Cheeke,
\textit{Fundamentals and Applications of Ultrasonic Waves}.
In CRC Series in Pure and Applied Physics, Editor-in-Chief D. Basu
(CRC Press, Boca Raton, 2002).

\bibitem{Dukhin02}
A. S. Dukhin and P. J. Goetz,
\textit{Ultrasound for Characterizing Colloids: Particle Sizing, Zeta Potential, Rheology}.
In series \textit{Studies in Interface Science}, Vol. 15, series editors D. M\"obius and R. Miller
(Elsevier, Amsterdam, 2002).

\bibitem{Kundu04}
\textit{Ultrasonic Nondestructive Evaluation: Engineering and Biological Material Characterization}.
Edited by T. Kun\-du (CRC Press, Boca Raton, 2004).

\bibitem{Hill04}
\textit{Physical Principles of Medical Ultrasonics}, 2nd ed.
Editors C. R. Hill, J. C. Bamber, and G. R. ter Haar
(Wiley, Chichester, 2004).

\bibitem{Rumack05}
\textit{Diagnostic Ultrasound}, 3rd ed.
Editors C. Rumack, S. Wilson, J. W. Charboneau, and J.-A. Johnson
(Elsevier Mosby, St. Louis, 2005).

\bibitem{Attwood81}
D. Attwood, L. Johansen, J. A. Tolley, and J. Rassing,
Int. J. Pharm. \textbf{9}, 285 (1981).

\bibitem{Tolley83}
J. A. Tolley and J. Rassing,
Int. J. Pharm. \textbf{14}, 223 (1983).

\bibitem{Tata93}
D. B. Tata, G. Hann, and F. Dunn,
Ultrasonics \textbf{31}, 447 (1993).

\bibitem{Ishihara93}
K. Ishihara,
J. Acoust. Soc. Am. \textbf{94}, 1176 (1993).

\bibitem{Jeffers95}
R. J. Jeffers,
J. Acoust. Soc. Am. \textbf{98}, 2380 (1995).

\bibitem{Tachibana98}
K. Tachibana and T. Uchida,
J. Acoust. Soc. Am. \textbf{103}, 2941 (1998).

\bibitem{Curra03}
F. P. Curra and L. A. Crum,
Acoust. Sci. Tech. \textbf{24}, 343 (2003).

\bibitem{Wu05}
J. Wu, J. Pepe, and M. Rincon,
J. Acoust. Soc. Am. \textbf{117}, 2473 (2005).

\bibitem{Gavish83}
B. Gavish, E. Gratton, and C. J. Hardy,
Proc. Nat. Acad. Sci. \textbf{80}, 750 (1983).

\bibitem{Gekko79}
K. Gekko and H. Noguchi,
\JPC \textbf{83}, 2706 (1979).

\bibitem{Gekko89}
K. Gekko and Y. Hasegawa,
\JPC \textbf{93}, 426 (1989).

\bibitem{Gekko91}
K. Gekko and K. Yamanagi,
J. Agric. Food Chem. \textbf{39}, 57 (1991).

\bibitem{Choi86}
P. K. Choi, J. R. Bae, and K. Takagi,
J. Acoust. Soc. Am. \textbf{80}, 1844 (1986).

\bibitem{Choi87}
P. K. Choi, J. R. Bae, and K. Takagi,
Jpn. J. Appl. Phys. Part 1 \textbf{26} Suppl., 32 (1996).

\bibitem{Bae96}
J.-R. Bae,
Jpn. J. Appl. Phys. Part 1 \textbf{35}, 2934 (1996).

\bibitem{Sakai00}
H. Sakai, K. Imai, M. Tanaka, M. Sonoyama, and S. Mitaku,
Jpn. J. Appl. Phys. Part 1 \textbf{39}, 2948 (2000).

\bibitem{Pethrick83}
R. A. Pethrcik,
Prog. Polym. Sci. \textbf{9}, 197 (1983).

\bibitem{Pavlovskaya92}
G. E. Pavlovskaya, D. J. McClements, and M. J. W. Povey,
Food Hydrocolloids \textbf{6}, 253 (1992).

\bibitem{Kharakoz89}
D. P. Kharakoz,
Biophys. Chem. \textbf{34}, 115 (1989).

\bibitem{Kharakoz91}
D. P. Kharakoz,
\JPC \textbf{95}, 5634 (1991).

\bibitem{Kharakoz93}
D. P. Kharakoz and A. P. Sarvazyan,
Biopolymers \textbf{33}, 11 (1993).

\bibitem{Ravichandran91}
G. Ravichandran, S. Adilakshmi, A. S. Rao, and T. K. Nambinarayanan,
Acustica \textbf{75}, 224 (1991).

\bibitem{Shin94}
D. O. Shin, E. J. Kim, and S. W. Yoon,
J. Acoust. Soc. Am. \textbf{96}, 3347 (1994).

\bibitem{Chalikian95}
T. V. Chalikian, V. S. Gindikin, and K. J. Breslauer,
J. Mol. Biol. \textbf{250}, 291 (1995).

\bibitem{Kitamura95}
H. Kitamura, B. Sigel, J. Machi, E. J. Feleppa, J. Sokilmelgar, A. Kalisz, and J. Justin,
Ultrasound in Medicine and Biology \textbf{21}, 827 (1995).

\bibitem{Karabutov98}
A. A. Karabutov and N. B. Podymova,
J. Acoust. Soc. Am. \textbf{103}, 3039 (1998).

\bibitem{Landau84}
L. D. Landau and E. M. Lifshitz,
\textit{Fluid Mechanics}.
In Landau and Lifshitz Course of Theoretical Physics, Vol. 6
(Pergamon Press, Oxford, 1984).

\bibitem{Richards39}
W. T. Richards,
\RMoP \textbf{11}, 36 (1939).

\bibitem{Markham51}
J. J. Markham, R. T. Beyer, and R. B. Lindsay,
\RMoP \textbf{23}, 353 (1951).

\bibitem{Bauer49}
E. Bauer,
Proc. Phys. Soc. \textbf{62A}, 141 (1949).

\bibitem{Gierer50a}
A. Gierer and K. Wirtz,
\PRev \textbf{79}, 906 (1950).

\bibitem{Teubner79}
M. Teubner,
\JPC \textbf{83}, 2917 (1979).

\bibitem{Bhattacharjee81}
J. K. Bhattacharjee and R. A. Ferrell,
\PRA \textbf{24}, 1643 (1981).

\bibitem{Narasimham90}
A. V. Narasimham,
Acustica \textbf{71}, 233 (1990).

\bibitem{Martynov01}
G. A. Martynov,
Theor. Math. Phys. \textbf{129}, 1428 (2001).

\bibitem{Delgado05}
A. V. Delgado, E. Gonz\'alez-Caballero, R. J. Hunter, L. K. Koopal, and J. Lyklema,
Pure Appl. Chem. \textbf{77}, 1753 (2005).

\bibitem{Hall47}
L. Hall,
\PRev \textbf{71}, 318 (1947).

\bibitem{Hall48}
L. Hall,
\PRev \textbf{73}, 775 (1948).

\bibitem{Hirata81}
F. Hirata and P. J. Rossky,
\CPL \textbf{83}, 329 (1981).

\bibitem{Hirata82}
F. Hirata, B. M. Pettitt, and P. J. Rossky,
\JCP \textbf{77}, 509 (1982).

\bibitem{Hirata83}
F. Hirata, P. J. Rossky, and B. M. Pettitt,
\JCP \textbf{78}, 4133 (1983).

\bibitem{Hirata03}
\textit{Molecular Theory of Solvation}, edited by F. Hirata.
In series Understanding Chemical Reactivity, Vol. 24,
series editor P. G. Mezey (Kluwer, Dordrecht, 2003).

\bibitem{Imai05}
T. Imai, R. Hiraoka, A. Kovalenko, and F. Hirata,
\JACS \textbf{127}, 15334 (2005).

\bibitem{Yoshida06}
N. Yoshida, S. Phongphanphanee, Y. Maruyama, T. Imai, and F. Hirata,
\JACS \textbf{128}, 12042 (2006).

\bibitem{Omelyan03}
I. Omelyan, A. Kovalenko, and F. Hirata,
J. Theor. Comput. Chem. \textbf{2}, 193 (2003).

\bibitem{Chong98}
S.-H. Chong and F. Hirata,
\JCP \textbf{108}, 7339 (1998).

\bibitem{Chong99a}
S.-H. Chong and F. Hirata,
\JCP \textbf{111}, 3083 (1999).

\bibitem{Chong99b}
S.-H. Chong and F. Hirata,
\JCP \textbf{111}, 3095 (1999).

\bibitem{Yamaguchi01}
T. Yamaguchi and F. Hirata,
\JCP \textbf{115}, 9340 (2001).

\bibitem{Yamaguchi03a}
T. Yamaguchi, S.-H. Chong, and F. Hirata,
\MP \textbf{101}, 1211 (2003).

\bibitem{Yamaguchi04b}
T. Yamaguchi, S.-H. Chong, and F. Hirata,
\JML \textbf{112}, 117 (2004).

\bibitem{Kobryn05}
A. E. Kobryn, T. Yamaguchi, and F. Hirata,
\JCP \textbf{122}, 184511 (2005).

\bibitem{Kobryn06}
A. E. Kobryn, T. Yamaguchi, and F. Hirata,
\JML \textbf{125}, 14 (2006).

\bibitem{Yamaguchi03b}
T. Yamaguchi, T. Matsuoka, and S. Koda,
\JCP \textbf{119}, 4437 (2003).

\bibitem{Zubarev74}
D. N. Zubarev,
\textit{Nonequilibrium Statistical Thermodynamics}
(Consultant Bureau, New York, 1974).

\bibitem{Kubo57a}
R. Kubo,
\JPSJ \textbf{12}, 570 (1957).

\bibitem{Kubo57b}
R. Kubo, M. Yokota, and S. Nakajima,
\JPSJ \textbf{12}, 1203 (1957).

\bibitem{Harp70}
G. D. Harp and B. J. Berne,
\PRA \textbf{2}, 975 (1970).

\bibitem{Berne70}
B. J. Berne and G. D. Harp,
\textit{On the calculation of time correlation functions.}
In Advances in Chemical Physics, Vol. XVII, p. 63-227.
Edited by I. Prigogine and S. Rice
(Intercsience Publishers, New York, 1970).

\bibitem{Berne76}
B. J. Berne and R. Pecora,
\textit{Dynamic Light Scattering. With Applications to Chemistry, Biology, and Physics}
(John Wiley \& Sons, New York, 1976)
[Reprinted by Dover Publications, New York, 2000].

\bibitem{Steele87}
W. A. Steele,
\MP \textbf{61}, 1031 (1987).

\bibitem{Mori65a}
H. Mori,
\PTP \textbf{33}, 423 (1965).

\bibitem{Mori65b}
H. Mori,
\PTP \textbf{34}, 399 (1965).

\bibitem{Balucani94}
U. Balucani and M. Zoppi, \textit{Dynamics of the Liquid State}.
In Oxford Series on Neutron Scattering in Condensed Matter, Vol. 10
(Clarendon Press, Oxford, 1994).

\bibitem{Balucani03}
U. Balucani, M. H. Lee, and V. Tognetti,
\PR \textbf{373}, 409 (2003).

\bibitem{Chong98b}
S.-H. Chong and F. Hirata,
\PRE \textbf{58}, 6188 (1998).

\bibitem{Chong02}
S.-H. Chong and W. G\"otze,
\PRE \textbf{65}, 41503 (2002).

\bibitem{Egelstaff92}
P. A. Egelstaff,
\textit{An Introduction to the Liquid State}, 2nd ed.
In Oxford Series on Neutron Scattering in Condensed Matter, Vol. 7
(Clarendon Press, Oxford, 1992).

\end{thebibliography}
\end{document}